# Drying of Bio-colloidal Sessile Droplets:

# Advances, Applications, and Perspectives

Anusuya Pal[a,b*] and Amalesh Gope[c] and Anupam Sengupta[d]

[a] Department of Physics, University of Warwick, Coventry, CV47AL, UK

[b] Department of Physics, Worcester Polytechnic Institute, MA, 01601, USA

[c] Department of Linguistics and Language Technology,

Tezpur University, Tezpur, Assam, 784028, India

[d] Physics of Living Matter, Department of Physics and Materials Science,

University of Luxembourg, 162 A, Avenue de la Faïencerie,

L-1511 Luxembourg City, Luxembourg

## Abstract

Drying of biologically-relevant sessile droplets, including passive systems such as DNA, proteins, plasma, and blood, as well as active microbial systems comprising bacterial and algal dispersions, has garnered considerable attention over the last decades. Distinct morphological patterns emerge when bio-colloids undergo evaporative drying, with significant potential in a wide range of biomedical applications, spanning bio-sensing, medical diagnostics, drug delivery, and antimicrobial resistance. Consequently, the prospects of novel and thrifty bio-medical toolkits based on drying bio-colloids have driven tremendous progress in the science of morphological patterns and advanced quantitative image-based analysis. This review presents a comprehensive overview of bio-colloidal droplets drying on solid substrates, focusing on the experimental progress during the last ten years. We provide a summary of the physical and material properties of relevant bio-colloids and link their native composition (constituent particles, solvent, and concentrations) to the patterns emerging due to drying. We specifically examined the drying patterns generated by passive bio-colloids (e.g., DNA, globular, fibrous, composite proteins, plasma, serum, blood, urine, tears, and saliva). This article highlights how the emerging morphological patterns are influenced by the nature of the biological entities and the solvent, micro- and global environ- mental conditions (temperature and relative humidity), and substrate attributes like wettability. Crucially, correlations between emergent patterns and the initial droplet compositions enable the detection of potential clinical abnormalities when compared with the patterns of drying droplets of healthy control samples, offering a blueprint for the diagnosis of the type and stage of a specific disease (or disorder). Recent experimental investigations of pattern formation in the bio-mimetic and salivary drying droplets in the context of COVID-19 are also presented. We further summarized the role of biologically active agents in the drying process, including bacteria, algae, spermatozoa, and nematodes, and discussed the coupling between self-propulsion and hydrodynamics during the drying process. We wrap up the review by highlighting the role of cross-scale in-situ experimental techniques for quantifying sub-micron to micro-scale features and the critical role of cross-disciplinary approaches (e.g., experimental and image processing techniques with machine learning algorithms) to quantify and predict the drying-induced features. We conclude the review with a perspective on the next generation of research and applications based on drying droplets, ultimately enabling innovative solutions and quantitative tools to investigate this exciting interface of physics, biology, data sciences, and machine learning.

* apal@wpi.edu

**HIGHLIGHTS**

1. Drying of sessile bio-colloidal droplets (biomolecules, -fluids, microbial active matter)

2. Catalogs and links dried patterns to clinical symptoms and diseases

3. Emerging cross-scale experimental techniques with machine learning algorithms

4. Perspectives on approaches for disease diagnosis, bio-sensing, and pathogenesis

**ABBREVIATIONS**

**List of abbreviations**

1,2-dihexadecanoyl-sn-glycero-3-phosphocholine (DPPC)

Acoustic Mechanical Impedance (AMI)

Antimicrobial Resistance (AMR)

Artificial Neural Networks (ANN)

Atomic Force Microscopy (AFM)

Blood Pattern Analysis (BPA)

Bovine Serum Albumin (BSA)

Complete Blood Count (CBC)

Computational Fluid Dynamics (CFD)

Confocal Fluorescence Microscopy (CFM)

Confocal Laser Scanning Microscopy (CLSM)

Confocal Microscopy (CM)

Convolution Neural Networks (CNN)

Critical Micellar Concentration (CMC)

Decision Tree (DT)

De-ionized (DI)

Deoxyribonucleic acid (DNA)

Drop Coating Deposition Raman Spectroscopy (DCDRS)

Dynamic Light Scattering (DLS)

First-order Statistics (FOS)

Fluorescence Microscopy (FM)

Fourier Transform Infrared (FTIR)

Gray Level Co-occurrence Matrix (GLCM)

Haematocrit (HCT)

Human Serum Albumin (HSA)

Infrared (IR)

K-Nearest Neighbors (KNN)

Liquid Crystals (LCs)

Logistic Regression (LR)

Lysozyme (Lys)

Machine Learning Algorithms (MLAs)

Mean Cell Volume (MCV)

Minimal Salts Medium (MSM)

Naive Bayes (NB)

Nanoparticles (NPs)

Neural Networks (NN)

Optical Microscopy (OM)

Partial least square (PLS)

Particle Image Velocimetry (PIV)

Polarizing Confocal Microscopy (PCM)

Polarizing Optical Microscopy (POM)

Principal Component Analysis (PCA)

Phosphate Buffer saline (PBS)

Raman Spectroscopy (RS)

Random Forest (RF)

Receiving Operating Curve (ROC)

Red Blood Cell (RBC)

Relative humidity (RH)

Ribonucleic acid (RNA)

Room Temperature (RT)

Scanning Electron Microscopy (SEM)

Support Vector Machine (SVM)

Surface-enhanced Raman scattering (SERS)

Transmission Electron Microscopy (TEM)

Virus Emulating Particles (VEP)

White Blood Cell (WBC)

## I.  INTRODUCTION

Pattern formation on solid surfaces due to evaporative drying of a droplet containing non-volatile particles is ubiquitous in a range of natural and synthetic settings. Non-volatile particles, commonly known as colloids, are omnipresent in droplets originating from different sources and have been studied for a long due to their fundamental and applied ramifications [1, 2]. The observations of drying sessile droplets are commonplace- both in a scientific lab and in nature. The spatio-temporal evolution of the drying patterns due to the changes in mass and concentration is a non-linear coupled phenomenon where multiple parameters, including environmental conditions [temperature and Relative Humidity (RH)], size and shape of inclusions, and their activity (namely, passive and active agents) mediate the formation of the dried morphology. Consequently, the field of drying droplets, specifically those with biological entities, has attracted immense multi-disciplinary attention over the last decades. For instance, there are several ways to phenomenologically interpret the drying of droplets: for a surface scientist, droplet drying is underpinned by the wetting and interfacial behavior (i.e., how the droplet interacts and wets the substrate) [3], while for a fluid physicist, the hydrodynamic flows and instabilities are of potential interest; and from a mechanical perspective, the solid mechanics of such an emergent system offers intriguing trade-offs among different stresses that underlie crack propagation and morphology [4].

Several studies conducted experiments to explore evaporative drying (not limited to sessile) and proposed different mathematical models [5, 6]. However, Deegan's landmark paper that proposed the famous 'coffee-ring effect' [7, 8] is often acknowledged as the first experimental evidence of the drying droplets where the physical mechanism (including mass transport and capillary flow) is explained in-depth. Looking further back, Robert Brown's pioneering work in the 80's on the motion of pollen particles in a fluid (thanks to Goldstein [9]), and the evaporation of fluid behavior – concerning solid particulates – was one of the earliest works on the drying of droplets [10, 11].

In simple terms, we can define colloids (also called colloidal solutions or colloidal systems) as minute particles dispersed in a medium, where the medium is a liquid (in most cases, aqueous medium, and will be the context of this review). The two essential properties of colloids are (a) the size of the particles ranges from 1 to 1000 nm (however, the micron-sized colloids have been extensively studied; for example, the size of White Blood Cell (WBC) in human blood is $\sim$15 $\mu$m, and (b) they do not sediment at the droplet-substrate base if left undisturbed. The bio-colloids are colloids that have biological relevance. This review divides these bio-colloids into two broad categories: (a) passive bio-colloids and (b) active bio-colloids. The active bio-colloids include microbes – bacteria, bacterial spores, bacteriophage, spermatozoa, phytoplankton (algae), and worms (nematodes). The passive bio-colloids are further divided into two sub-categories: (i) biomolecules and (ii) bio-fluids. The term 'biomolecule' is loosely used to refer to the bio-colloid present in mammals (animals or humans), essentially for different biological processes. It includes DNA and globular proteins [such as Lysozyme (Lys) and Bovine Serum Albumin (BSA)]; fibrous proteins (such as collagen), and composite proteins (where a protein cannot be strictly divided into the globular or fibrous category); such as microtubules, casein, and whey proteins. The term 'bio-fluid' is used for those bio-colloids which are excreted (urine, sweat, and saliva); secreted (breast milk, bile, tear); obtained with a needle (blood or cerebro-spinal fluid), or develop as a result of pathological circumstances (blister or cyst fluid). In this review, we will limit ourselves to urine, saliva, tear, and whole human blood (that can be centrifuged into plasma and blood serum).

Experimentally, the drying of colloidal droplets involves a relatively simple protocol. The colloidal system is prepared by adding different amounts of colloidal particles into the solvent, such as water or buffer. A substrate is chosen, on which a fixed volume of solution is pipetted as a droplet. The word 'sessile' qualifies droplets to rest on a supporting substrate in equilibrium (with respect to the gravity force). When the solvent evaporates from the droplet, the local concentration of colloidal particles increases as the solvent escapes from the system, ultimately resulting in a unique pattern (deposit) as the droplet dries completely. This signature pattern can be used as a fingerprint for the colloidal solution. In most cases, the droplet is not pinned to the substrate without such colloidal particles. The droplet radius shrinks as the droplet dries, and finally, no deposit resides. In other words, the higher the impurity (components) in the colloidal droplet, the more complex the residual pattern is once the droplet dries up. These emerging patterns play an essential role in many industrial processes, ranging from inkjet printing [12], film coating [13], and DNA microarrays [14] to disease diagnosis [15, 16].

In the context of bio-fluids, the first step in the protocol might require anti-coagulants and/or purification steps such as centrifugation to remove biological debris, including cells and tissue remnants. In active bio-colloidal microbial systems [17–19], typically, species-specific methods are required to grow the relevant species and prepare appropriate buffers for running the experiments. Additional parameters, including interfacial entrapment, physiological growth stages, external cues, and intrinsic stress levels, are experimentally relevant since these factors directly control the activity of the microbial systems [20–25], and, thus modulate the emergent drying patterns. Furthermore, underlying substrates play a crucial role in droplet fate, depending on their surface energy (wetting vs. non-wetting) and topography. To vary the drying conditions, the droplet on the substrate must be in a controlled chamber with controllable temperature and/or RH to increase or decrease the drying rate. Multiple studies have indicated that the drying rate leaves an imprint on the patterns [26–30], not only for biologically relevant systems but also in the case of simple colloidal solutions (such as polymers, nanoparticles, etc.). Typically, the droplets drying under an elevated temperature $(>$ Room Temperature $(\text{RT}) \sim 26^{\cdot}\text{C})$ impact the emergent morphological patterns, as the colloidal particles do not get enough time to segregate during the drying process [31, 32]. Thus, the spontaneous involvement of multiple factors during the drying process makes the underlying principles challenging to comprehend. Not surprisingly, it becomes more complex when (a) the number of components increases from one [33] to three [34] or more [35], and (b) these components further interact and self-assemble to form different structures [36].

There has been a significant effort in reviewing drying droplets using experiments [4, 15, 16, 37–44], theory, and computation [45, 46] over the last few years. The literature has also been reviewed for both experimental and numerical studies, focusing on the suppression of and the implication of the coffee-ring effect [47–49]. In recent years, the focus has been shifted to the progressive drying of ferrofluids and their morphological patterns in magnetic fields, various parameters of the magnetic nanoparticles (type, size, shape, charge, etc.), the surfactant properties, and the temperature and initial concentration. The results discuss the wetting, interfacial dynamics, surface stress, and hydrodynamic instabilities of drying ferrofluid systems [50].

Despite the numerous review articles on drying droplets, to our knowledge, drying bio-colloidal droplets– which entail cross-disciplinary and cross-scale approaches– have seldom been systematically reviewed. Such a review will be central to understanding bio-colloidal droplets' fate and their potential application in connecting patterns to biomedical disorders. The current review attempts to bridge the gap by providing a thorough review of sessile bio-colloidal droplets, summarizes the recent advances in the drying community, and finally, and provides new perspectives on this field. The bio-colloidal systems that are discussed in this review are (a) biomolecules that include DNA and various types of proteins (globular and fibrous, and composites), (b) complex naturally occurring bio-fluids (blood serum, plasma, and blood), and (c) microbial dispersion. The highlights of this review article are (a) a hierarchical description of the bio-colloids where the system complexity varies progressively from biomolecules to bio-fluids to microbes, (b) provide a catalog of the patterns in relation to the diseases and disorders with collating images, thereby providing a visual perspective to the readers, (c) drying of droplets of active microbial agents, and (d) a description of existing techniques and recent cross-scale approaches to understanding pattern formation in drying droplets.

As presented in Figure 1, the review spans different bio-colloidal systems (top-half), and recent progress in the fundamental and application (bottom-half) domains. The re-view is organized as follows: first, we will briefly discuss the passive and active bio-colloids (biomolecules to bio-fluids to microbes). Then, we will focus on recent experimental work on drying passive bio-colloids. The drying droplets and dried morphological patterns of microbial fluids are also examined. The cross-scale investigations involving emerging ex-perimental techniques, image processing tools, and machine learning algorithms on drying droplets are elaborated, offering new avenues to be used for bio-colloidal studies. Finally, we provide perspectives and an outlook on the fundamental understanding of bio-colloidal drying droplets, using bio-fluids as a tool for pre-diagnosis, biosensing, and pharmaceutical applications. In addition, perspectives on emergent virulence, pathogenesis, and implication on the anti-microbial resistance in the contribution of microbial drying droplets are also presented.

The relevant resources, such as reviews and journal articles, conference proceedings, book chapters, books, etc., were collected from the scientific databases (used PubMed, Elsevier, and Google scholar) using the search terms: 'drying drop', 'dried drop', 'evaporating drop', 'evaporation', 'protein drying', 'blood drying', 'crack patterns', 'morphology', 'diagnosis', etc. Additionally, the reference lists given in the resources were searched manually for further relevant publications. Only publications in English and published papers have been considered here. For this review article, our motivation is to summarize the cross-disciplinary and cross-scale line of research on drying bio-colloidal droplets in the sessile configuration. Results on polymer-based samples have been significantly reviewed; therefore, we did not include them in this article, and details could be found elsewhere [4, 39]. In this review article, we have primarily targeted the studies published in the last ten years. Noteworthy, we will use 'droplet' throughout this article to benefit the new researchers in this field; however, it must be noted that a droplet is often < 50 $\mu$m in diameter, and a drop is formed when the diameter becomes ~500 $\mu$m or bigger [4]. The fundamental concepts involving numerical studies of drying droplets are kept to a minimum and can be found in Ref [46]. We primarily focus on the two vital theoretical concepts relevant to the review: (a) the General physics of drying droplets and (b) Mass transport and pattern formation.

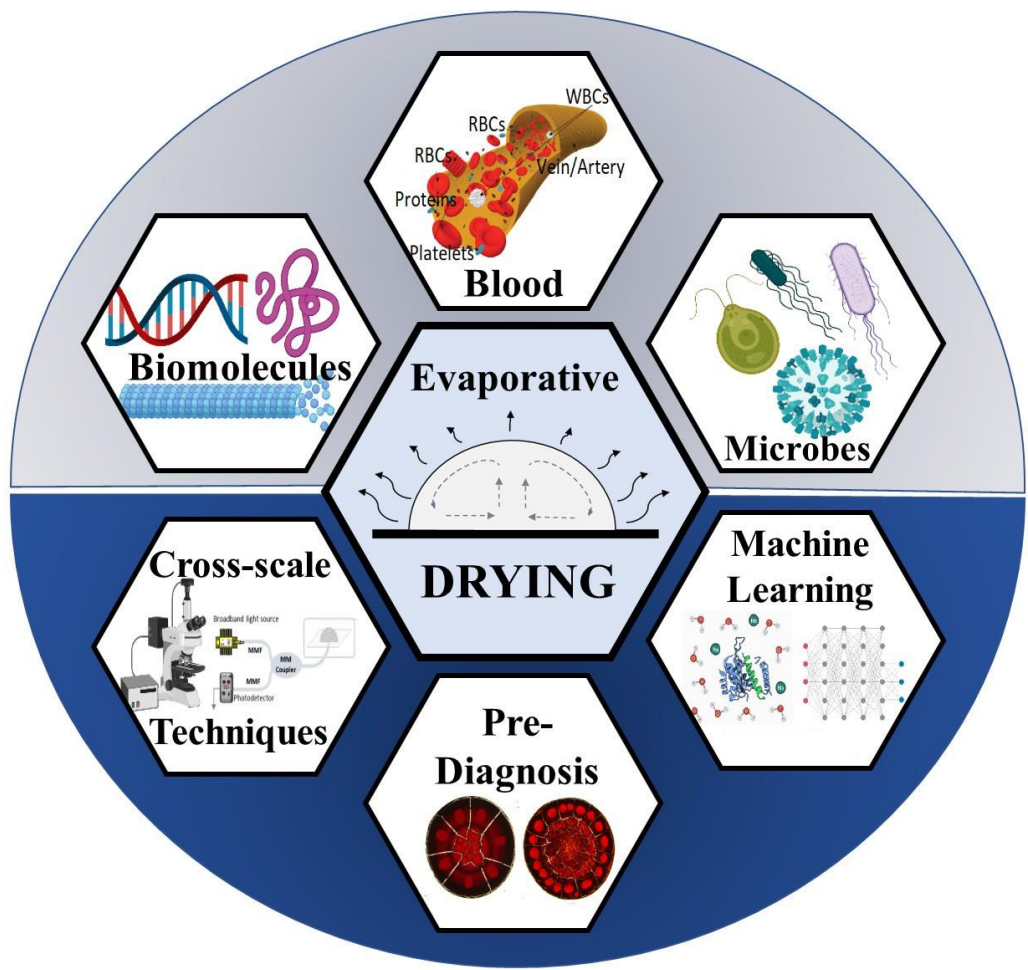

FIG. 1. Presents a visual depiction of the bio-colloidal systems reviewed in this article.

## A.    General physics of drying droplets

When a droplet of any bio-colloidal fluid solution is deposited on a substrate, the constituent particles are dispersed uniformly in the droplet, defining an initial equilibrium state. As the solvent evaporates from the droplet, the system is driven out of equilibrium, and gradients in concentration emerge, which generates different flows within the droplet. This process acts as an engine and continues until the stored chemical potential of the solvent is exhausted as the system relaxes to another equilibrium-like state but with the suspended colloids organized (or assembled) onto the substrate. Specifically, for the partial wetting case, primarily seen in the hydrophilic substrates, the droplet forms a spherical-cap shape geometry, and the solvent mass loss near the periphery is larger than the center. Different flows carry and transport the suspended particles within the droplet. Suppose the droplet is pinned to the substrate (which is the case most of the time for a bio-colloidal solution). In that case, the droplet radius is constant throughout the drying process and experiences increasing mechanical stress as the solvent tries to leave. And finally, as the last of the solvent evaporates and the droplet fractures to relieve the stresses, a macroscopic pattern (fingerprint) emerges. When these droplets are dried under uniform conditions (shape, size, surface, temperature, humidity, etc.), the final pattern appears to be a unique fingerprint dependent on the initial state of that bio-colloidal solution.

It is worth mentioning that different drying modes occur; for example, (a) Constant Contact Radius (*CCR*) mode, in which only the contact angle reduces as time progresses. The diameter of the droplets (or the contact radius) remains constant during the drying process. (b) Constant Contact Angle (*CCA*) mode when the contact radius of the droplet reduces with drying time without changing its contact angle. The simultaneous movement of the contact angle and the contact radius during the evolution of the drying process is called the Mixed Mode (*MM* ) [4]. Since the bio-colloidal droplets get pinned to the substrate, implying that the contact radius cannot shrink with drying time. In general, the bio-colloidal droplets show *CCR* mode. However, it is noticed that a 'fluid front' moves from the periphery to the central region despite having a constant contact radius. Different terminologies can be found for this 'fluid front'. Sometimes, it is called a 'sol-gel' front, 'transition front', or 'gelation front' depending on the system. Therefore, the terminology is contextual, and the reader should pay attention while defining it [51].

## B.  Mass transport and pattern formation

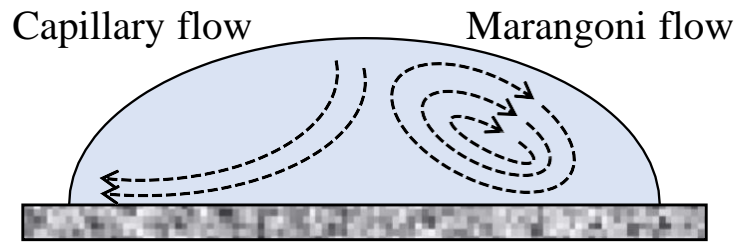

FIG. 2. Presents a schematic representation of distinct flow regimes that develop during a droplet's drying on a solid substrate. This illustration is adapted from [51].

Different types of convective flows occur in bio-colloidal systems. One type of convection is 'capillary convection', where a gradient of the Laplace pressure generates when there is any change (distortion) in the droplet shape. This convection can generate both inward and outward flow inside the droplet. The 'coffee-ring' effect is a well-known physical mechanism in a drying sessile colloidal droplet [7]. It shows that the evaporation rate is the highest near the periphery due to the curvature of the droplet. The colloidal particles present in the droplet are transported towards the periphery by the outward radial flow [also known as 'capillary convection' [see Figure 2)] to compensate for the excessive mass-loss. This transport will deposit more particles near the periphery than the central region. This deposition of the particles will develop a ring, called the 'coffee-ring' effect.

Capillary convection can also originate from the surface tension gradient along the droplet's surface. This is known as 'Bénard–Marangoni convection' (or Marangoni flow). This surface tension gradient can be induced by either temperature (referred to as 'thermal Marangoni') or the solute concentration (referred to as 'solutal Marangoni'). Besides these flows, interfacial flows are observed in the drying sessile droplets, especially for the hydrophobic surfaces. Due to the high contact angles, the evaporation rate is higher at the droplet apex compared to its periphery. The particles at the air-liquid interface prefer to move from the apex (higher curvature) to the periphery (lower curvature), and the particles accumulate near the periphery [52]. Differential curvatures can have significant implications in biological systems [53]; however, their role in drying bio-colloidal droplets remains to be explored. Another interesting study argues that the diffusion (here, non-directional bulk flow for mass transport), advection (here, horizontal mass transport towards the periphery), and capillary attraction (the ordered arrangement of the particles mainly induced by the capillary forces near the three-phase interface) are responsible for the emerging patterns in the drying droplets that are partially-wet to the substrate [54]. Often a combination of Marangoni and Capillary flows occur within a droplet. For instance, the Capillary flow is typically much stronger than the Marangoni flow in droplet drying at RT. In contrast, Marangoni flow is dominant for droplet drying on a heated substrate.

Various interactions (interactive and repulsive forces) between the substrate and the suspended particles are possible, known as DLVO interactions [55]; however, in the case of bio-colloids, interactions between particles also play a role. These interactions can be tuned by varying the suspension's pH (ionic behavior of these particles), substrate wetting properties from hydrophilicity to hydrophobicity, etc. Therefore, different interactions dictate the nature of the flows within a droplet, which further influences the kind of deposits the droplet will form once the solvent leaves the system. In this context, the residue left behind in the colloidal system is known as a 'pattern'. It varies from uniform, coffee-ring, dot-like to stick-slip patterns [40]. Various types of crack patterns also form, ranging from chaotic to ordered [34] to spiral cracks [33].

## II.   BIO-COLLOIDS WITHIN DRYING DROPLETS

This section briefly describes the physical and material properties of both passive and active bio-colloids, which have been studied as drying droplets. This information is essential as it sets a baseline to map the resulting drying-induced morphological patterns to the native (initial) states of the constituent particles. Nonetheless, this section does not cover all properties but the critical information relevant to the drying process. The targeted bio-colloids are biomolecules, bio-fluids, and microbial fluids.

### A. Biomolecules

*1. DNA– Genetic information carrier*

DNA is a polymer composed of two polynucleotide chains to form a double helix. It contains many genes, but not all genes get expressed simultaneously. Therefore, recent advancements focus on understanding gene delivery, gene profiling, and the self-assembly, microarray, and patterning of the DNA in different microenvironments via drying droplets and microfluidic techniques [14, 56]. Analysis of gene expression is essential as it paves the way to differentiate and compare healthy and diseased cell profiles. The DNA strand has two ends where the strand is made from alternating phosphate and sugar groups. A phosphate group is attached to the 5' carbon of a ribose (the 5' phosphoryl) at one end of the DNA strand. A hydroxyl group is attached to the 3' carbon of a ribose (the 3' hydroxyl) at the other end of the DNA strand. Therefore, their orientations along the sugar-phosphate backbone provide a directionality or polarity to each DNA strand. This promotes DNA to behave as lyotropic LCs, implying that their birefringence properties can be tuned by the increase or decrease of the initial concentrations [57]. It is noted that DNA is negatively charged– the bond between oxygen and phosphorus atoms is negative, making the phosphate backbone negative, hence the overall DNA structure negative [14, 58].

*2. Proteins: globular and fibrous proteins*

We broadly divide the protein-drying droplets into globular, fibrous, and composite proteins (a combination of globular and fibrous proteins). The differences between the globular and fibrous proteins are tabulated in Table I. It is observed that many sessile drying droplet studies have been performed on Lys and BSA. It is presumably due to ease of availability and lost cost; and because it is readily available in human fluids. For example, Lys is abundant in mucosal secretions, including tears, saliva, human milk, etc. A significant amount of Lys is also found in the chicken egg-white. Interestingly, BSA (globular albumin protein present in cow blood) is chemically identical to the globular albumin protein present in human blood [i.e., Human Serum Albumin (HSA)] [59]; however, HSA is much costlier than BSA. Though Lys and BSA show a globular nature, their physical characteristics are unique, as tabulated in Table II.

TABLE I: Difference between globular and fibrous proteins' physical characteristics.

| Characteristics | Globular Proteins | Fibrous Proteins |
|---|---|---|
| Shape | These are compact and spherical and folded in nature | These are elongated and have helical or sheet-like structures |
| Structure | Most of their hydrophobic residues are buried inside, and charged residues are at the surface | Each site of these proteins is exposed to interact with other particles |
| Functions | Mainly involved in the metabolic functions | Mainly constitute structural element |
| Solubility | Mostly soluble in water | Mostly insoluble in water |
| Sensitivity | Highly sensitive to changes in pH, temperature, etc. | Less sensitive to changes in pH, temperature, etc. |
| Example | Lys, BSA, HSA, haemoglobin | Actin, myosin, collagen, keratin |

TABLE II: Differences between Lys and BSA globular proteins' physical characteristics.

| Characteristics | Lys | BSA |
|---|---|---|
| Molecular mass | Each Lys has a molecular mass of 14.3 kDa | Each BSA has a molecular mass of 66.5 kDa |
| Shape | Roughly spherical shape | Ellipsoid shape |
| Dimension | $3.0 \times 3.0 \times 4.5$ nm$^3$ | $4.0 \times 4.0 \times 14.0$ nm$^3$ |
| Amino acids | 129 amino acids in a single polypeptide chain | 581 amino acids in a single polypeptide chain |
| Isoelectric point | 11.1 allows it to carry a net positive charge at a pH of 7 | 4.7 allows it to carry a net negative charge at a pH of 7 |
| Disulfide bridges | Lys has 4 bridges | BSA has 17 bridges |
| Denaturing temperature | $\sim 75\,^{\circ}$C | $\sim 65\,^{\circ}$C |

| | Lys is found in human mucosal secretions such as saliva, tears, etc., and raw egg-white solution | BSA predominantly acts as a transporter protein in the circulatory system |
|---|---|---|
| Predominance | | |

In contrast to these globular proteins, investigations have also been expanded to the fibrous proteins. For example, collagen is the structural and most abundant protein in the extracellular matrix found in animal bodies. It is the main component of the connective tissues, making up 25-35% of the whole-body protein content. The material properties of collagen vary; for example, these can be rigid (bone), compliant (tendon), and/or have a gradient from rigid to compliant (cartilage). Unlike other globular proteins, collagen has birefringence properties as many other structural proteins. These properties are typically studied under crossed-polarizing microscopy, where the analyzer and polarizer are set in perpendicular directions. Their structure, distribution, and alignment could be tuned by driving the system during drying. The substrate temperature change and the addition of probes (at the smaller concentrations for various ionic strengths) trigger the birefringence properties. Additionally, the disorder-to-order transition of their assembly leads to distinct pattern formation, which will be elaborated on in the next section.

Another category of biomolecules is based on the combination of both globular and fibrous proteins. We will limit ourselves to those investigated as sessile drying droplets: whey, caseins, and microtubules. The milk proteins are attractive for sessile-drying droplet configurations due to the different morphological signatures they leave behind [60]. Milk contains two main types of proteins– casein and whey. Casein represents $\sim$80%, while whey contributes $\sim$20% of the milk proteins. The molecular weight of casein is 19-25 kDa. The isoelectric point is 4.6. Unlike Lys and BSA, caseins contain no disulfide bonds. This also lacks a tertiary structure that makes it stable against heat denaturation.

Interestingly, caseins are conjugated proteins and can form a micelle. The micelle is relatively porous and is reactive to other substances. The casein micelles are 100−300 nm and exhibit a deformable sponge-like structure [61, 62]. Its stability depends on various factors: (a) the salt content, (b) pH, (c) temperature, and (d) dehydration rate. It implies that dehydration due to the drying process could impact their stability and change their native states over time. However, this is not captured just yet how the stability of these proteins decreases as the drying time increases. In contrast, the proteins appearing in the milk supernatant after precipitation at pH 4.6 are collectively called whey proteins. The average diameter of whey protein is 10–30 nm. Whey is a rigid globular protein, unlike casein, and can be considered a hard colloidal particle [63]. The isoelectric point is 3.5-5.2. It is more water-soluble than caseins and is subject to heat denaturation. Whey protein is again composed of two globular proteins: (a) 70% $\beta$-lactoglobulin, mostly in dimer form with the tertiary structure of molecular weight $\sim$36.6 kDa, and (b) 20% $\alpha$-lactalbumin with molecular weight 14.2 kDa [64]. The structural difference, such as spongy vs. hardness, and the weight difference within the whey protein lead to different mechanical instabilities during the drying process, which will be elaborated on in the next section.

Microtubules, composed of a single type of globular protein called tubulin, and different filamentous structures, are typically cylindrical asymmetric structures. These are composed of $\alpha$- and $\beta$-tubulin heterodimers where each monomer is a 55 kDa protein. These proteins are arranged in a head-to-tail fashion, leading to a polarity. Similar to collagen, these proteins possess birefringence properties, and their alignment could also be tuned by increasing the temperature of the solutions during the drying process. However, there is no causality between birefringence and temperature tunability. These usually vary in length but have a diameter of $\sim$25 nm. The dependence of the tubulin length as a function of drying rate has not been explored yet, but might give rise to different drying-induced morphological patterns. The diameter of the central hollow core is $\sim$15 nm, and the wall is $\sim$5 nm thick [65].

It is important to note that these proteins or protein solutions are not that simple and can be considered complex fluids in rheology. For instance, the initial concentration of the protein in the fluid influences its rheological characteristics. The same protein at low concentrations behaves as a Newtonian fluid, whereas it becomes non-Newtonian at high concentrations. Therefore, investigating the viscosity as a function of the protein's initial concentration and during the drying process is an essential missing link; however, such experimental studies are rare.

### B.  Human blood, plasma, and associated components

In a typical lab setting, body fluids (such as blood, saliva, urine, etc.) are used as a biomarker to diagnose a disease or disorder. In terms of pathological tests, human blood is an invasive specimen, i.e., a medical instrument enters the body (i.e., a needle is needed to take out the blood). At the same time, urine and saliva are non-invasive specimens (do not require breaking the skin or entering the body). Blood is collected and used for most diagnoses of diseases. It is a fluid consisting of plasma (composed of ∼93% water-carrying ions, nutrients, and ∼7% proteins) and associated cellular components [Red Blood Cells (RBCs) or erythrocytes, WBCs or leukocytes, platelets, or thrombocytes]. Different types of WBCs, such as monocyte, neutrophils, basophils, and echinophils, are present in our blood, making it a naturally occurring multi-component system [see Figure 3(I-II)]. A 1 $\mu L$ of whole blood contains 400 to $500 \times 10^4$ of RBCs, 0.5 to $1 \times 10^4$ of WBCs, 14 to $40 \times 10^4$ of platelets depending on the pathological condition of a donor, along with a small quantity of plasma proteins (fibrinogen, immunoglobulins, albumin) and salt ions [66]. Therefore, if the blood is diluted with water, the relative concentrations remain the same, whereas the overall mechanism of particle-particle interactions are expected to change, more details related to the drying dynamics can be found in the next section.

RBCs contain hemoglobin and a membrane as well. A membrane is made of 19.5% water, 39.6% proteins, 35.1% lipids, and 5.8% (w/w) carbohydrates [51, 67]. With no surprise, one can understand why there is bio-mimetic saliva and urine, but we have not yet prepared the bio-mimetic blood. It is to be noted that the term 'whole' term signifies that the human whole blood consists of both plasma and cellular components. However, the whole blood could separate into parts if kept undisturbed in the test tubes due to the difference in densities. The standard method used for the *in-vitro* experiments is centrifugation. The coagulation process can be prevented, and all components are suspended when an anticoagulant is added to the blood immediately after collection. If this anti-coagulant added blood is undisturbed, the heavier cells (mostly RBCs) will sink to the bottom. The upper layer containing plasma includes all the components without the cells. Under centrifugation, a layer of WBCs called the buffy coat forms, separating the plasma from RBCs. This plasma retains fibrinogen. On the other hand, without an anti-coagulant, the fibrinogen as a clotting agent efficiently removes RBCs from the plasma as a solid mass. The serum will then form containing plasma without fibrinogen and is most widely used in research [see Figure 3(I-II)]. Whether there is any significant relation between the type of anticoagulant and the pattern formation in the blood-drying droplets remains an open question. However, one can expect an additional influence on the pattern generated from the drying droplets

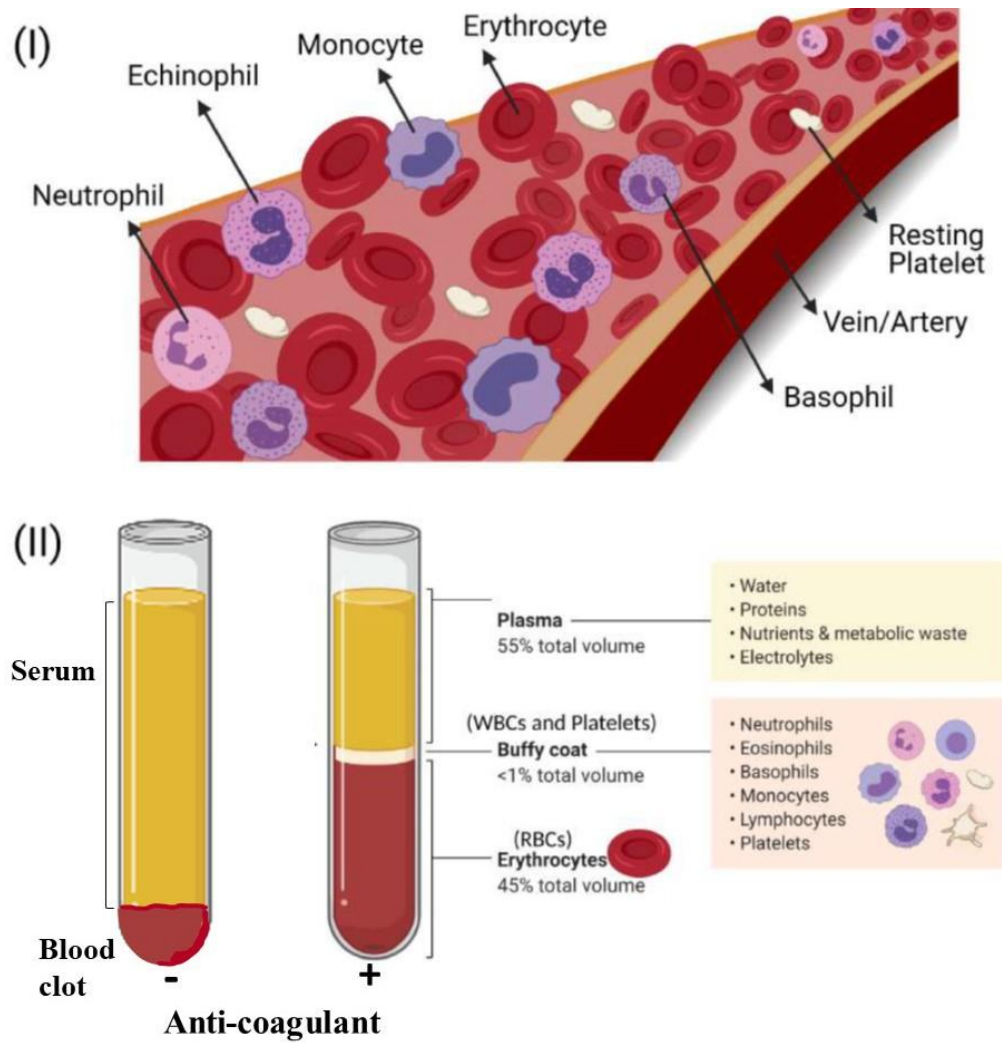

FIG. 3. (I) Cellular components (RBCs, WBCs, and platelets) of the human blood transported through the blood vessel (artery or vein). Various types of WBCs (monocyte, echinophil, neutrophil, basophil) are also shown. (II) Illustration of the anticoagulant added blood that could be centrifuged into different layers, including plasma, a buffy white coat, and RBCs. In the absence of an anti-coagulant, the blood clots and settles in the bottom of the tube. The serum is formed in the upper part of the tube. The images are adapted from [51]

.

due to the compositional variability and the preparation procedure (e.g., centrifugation and addition of anticoagulants). In a few instances, the term 'haematocrit' (HCT) is used. It estimates the concentration of blood cells in the plasma and is measured as a ratio of the packed cells to the total blood volume. Furthermore, the volume of RBCs in Complete

Blood Count (CBC) is measured as Mean Cell Volume (MCV). Furthermore, the bilirubin concentration (a yellowish pigment) is also checked while testing CBC. High bilirubin levels mean that either RBCs are breaking down at an unusual rate or that the liver is not breaking down waste properly and clearing the bilirubin from your blood– a coupon sign of jaundice, anemia, and liver disease [68].

The typical size of RBCs, WBCs, and platelets are 6-8 $\mu$m, $\sim$15 $\mu$m, and 2-3 $\mu$m, respectively [69]. It is essential to know that these associated cellular components in the blood can alter their shapes in response to the toxicity, fluid's ion concentration, pH (hypertonic or hypotonic environment), shear, and mechanical stresses during the drying process. The process may lead to different self-assembling mechanisms, deposition, and unique pattern formation. More so, these stresses influence the biological activity of such components. For example, the healthy RBCs are of bi-concave or discoid shape. However, they transform into different forms such as ellipsoidal, echinocyte (a round shape with short blunt spicules), sickle (crescent moon), teardrop, etc. [67]. For the *in-vitro* experiments, it is found that the transformation of discoid to echinocyte shape could be due to the contact of the RBCs with substrates like glass or plastics. The reason for this change is not yet known; however, one could argue the reason as the electrostatic interactions. In the *in-vivo* case, such a variation is observed when the person undergoes a cardiopulmonary bypass or decompression phase [70]. On the other hand, the WBCs are white in color and generally contain irregular and nucleated structures. These structures can transform into different functional forms, irregular troughs, and ridges (microvilli structures) [71, 72]. The platelet follows a discoid and an anuclear structure when inactivated (or in the resting phase). However, it changes into its spread form with extended filopodia on its activation [see Figure 3(I-II)]. The cytoskeleton of platelets is composed of actin and actin-binding proteins; these can polymerize and activate platelets in response to environmental or chemical signals [73].

### C.  Microbes as active bio-colloids

So far, we have discussed the physical properties of colloidal particles such as biomolecules, human blood, etc., which are essentially passive in nature. The interpretation becomes far more complicated when the underlying mechanisms and dynamics of the active bio-colloids are considered. In addition to the short-range interactions, such as van der Waals, electro-static interactions, etc., that are commonly found in the passive systems, the hydrodynamic interactions, substrate-mediated cross-interactions, etc., influence the collective behavior, motility, and hierarchical structure formation of the active systems [74]. In particular, the complexity arises due to two simultaneous processes: (a) consumption of energy (either stored or externally as nutrients) driven by microscopic entities (or motors), and (b) generation of forces in directions related to the flow of energy, which is rarely found outside the living world [20]. For instance, communities comprising populations of various microbial species are of centimeter-scale, the populations of single species are of millimeter-scale, the individual cells are of micrometer scale, and the cellular processes such as phase separation, organelle compartmentalization, etc., are typically in the micrometer scale [20].

Drying droplet experiments have been performed on different types of bacteria such as *Escherichia coli*, *Pseudomonas aeruginosa*, *Salmonella typhimurium*, *Staphylococcus epidermidis*, *Bacillus subtilis*, *Latilactobacillus sakei*. The Gram staining method of bacterial differentiation identifies two classes of bacteria: (a) Gram-positive (have a thick peptidoglycan layer in the cell wall) and (b) Gram-negative (have a thin peptidoglycan layer). The name 'Gram-positive' is given as they retain the crystal violet stain used in this differentiation method. The cell wall envelopes the cell membrane, mainly protecting the cell from changes in the micro-environment (osmotic pressure, pH, etc.). Though the presence of the cell wall affects many intrinsic characteristics of these bacteria, it is not yet explored if there is any significant influence of the cell wall properties on the pattern formation. *E. coli* cells are prokaryotes, and of size $\sim 1\ \mu m$, with a flagellum of length 5-10 $\mu m$ [75], whereas the eukaryotic microorganism, *C. reinhardtii* is $\sim 10\ \mu m$ in size, with flagella length of $\sim 10\ \mu m$ [76]. Some strains (CC-125 and CC-124) of *C. reinhardtii* also contain an eyespot— a light-sensitive organelle due to which they can perform phototaxis. The primary function of a flagellum is the movement of the cell (locomotion). Still, it serves a sensory function, sensitive to mechanical and chemical boundaries, temperature, etc. Interestingly, the motility of any microorganism (irrespective of prokaryotic and eukaryotic) affects the drying dynamics. Furthermore, unique patterns emerge when the phototaxis of *C. reinhardtii* is investigated during the drying process. This indicates that the drying process drives the system so that we can study the physiological states of these microbes by relating the resulting morphological patterns to the initial conditions of these active colloids.

It is to be noted that the history of sample preparation, type of microorganism, culture,

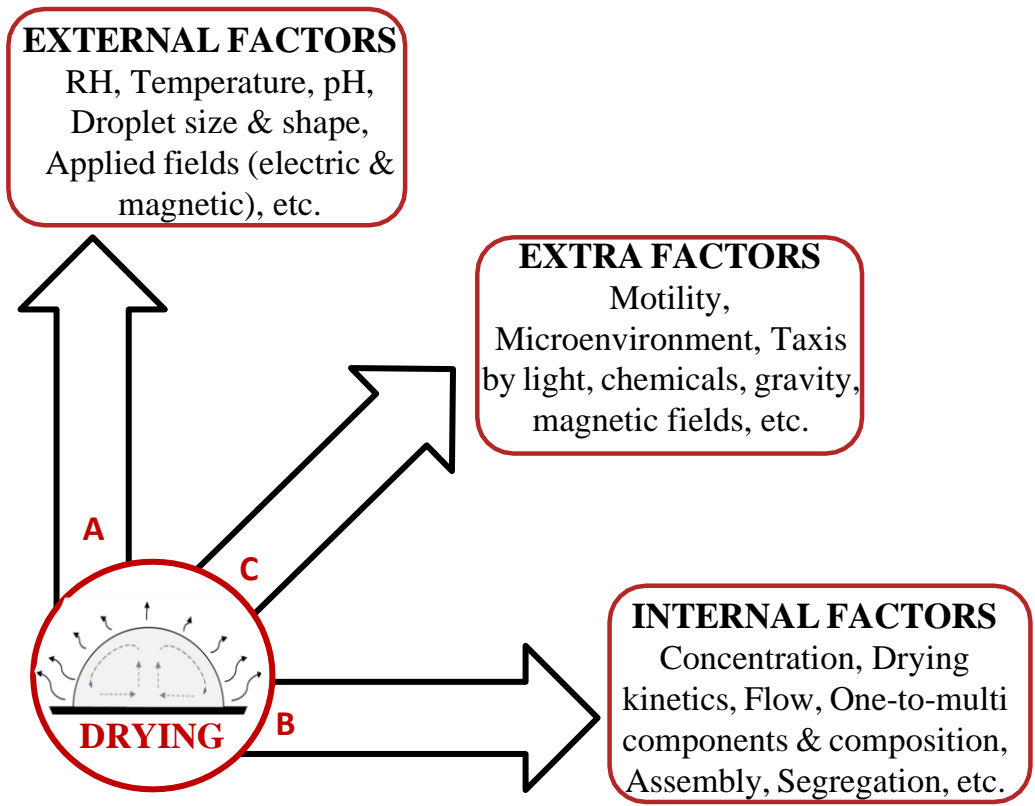

FIG. 4. Key parameters parameters influencing the drying experiments in bio-colloidal droplets: A-axis and Baxis are relevant for all passive bio-colloids. For example, we can see how one-to- multi components in passive bio-colloids affect pattern formation [35]. The substrate temperature affecting the drying rate is present in the passive bio-colloidal droplets [31] and in the active bio-colloids [19]. In addition, C-axis is particularly for the experiments with the active bio-colloids; for instance, how the chemotaxis affects the drying pattern of bacteria droplets [77].

growth conditions, and optimal conditions are crucial in understanding microbial pattern formation. In a nutshell, we can say that the external and internal factors are considered for any experiments and/or finding any insights. In contrast, some extra factors become essential if anyone wants to perform drying droplet experiments with the active particles. The key variables are listed as a schematic in Figure 4. It is also found that the microorganisms tend to secrete different polymer and lipid-like substances as these replicate. This secretion changes the fluid properties (in a colloidal sense—more viscous), affecting the dried morphological patterns. To understand what is the actual influence of the secreted material, one needs to perform all the experiments (a) in the fresh buffer by centrifuging the microor-ganisms and discarding the buffer as supernatant and/or, (b) using that supernatant buffer and performing the drying experiments with the polymer beads of similar sizes. Another important thing is to note that it is better to check the growth curve and see if these microorganisms properly replicate whenever one has to investigate the influence of motility during the drying process. Another point has to keep in mind that these organisms tend to adhere to the glass surface if an untreated glass substrate is used.

To solve this, most of the time, during any bio-physical experiments, the glass slides are covered by dilute BSA solution; however, for the drying experiments, typically, it is not recommended to do so. It is because the emerging patterns might get influenced by this added BSA. Therefore, this stickiness should be handled differently; for instance, we can treat the glass slides and change their wettability to be a little hydrophobic.

## III.    A HIERARCHICAL FRAMEWORK OF BIO-COLLOIDAL DROPLETS

### A.    Drying droplets of DNA solutions

One crucial category of biomolecules is Deoxyribonucleic acid (DNA), Ribonucleic acid (RNA), and nucleic acids. A vast body of literature has been available since the last decade [56, 57, 78, 79], signifying the progress in drying droplets. In this review, we will only highlight the recent progress on drying droplets of DNA. From the outcome of each research, it is clear that the morphological patterns depend on multiple parameters such as concentration, substrate type (wettability), drying conditions, etc. Interestingly, it is also shown how the drying droplet pattern can be leveraged for DNA detection using hybridization-induced suppression of the coffee ring effect [80].

In the last decade, drying patterns of DNA solution in the presence of colloidal particles (polymers, NPs) became a focal point of research due to its rich physics and relevance for potential applications. The DNA-colloidal system was also used to examine drying-induced interfacial hydrodynamics. The self-assembled morphology emerges due to competing interactions between intermolecular and interfacial effects [79, 83]. Suspensions with high DNA but initially low colloidal concentrations favored the formation of a multiple-ring pattern. The colloidal particles' size also impacted the dried morphology yielding curtain-like periodic patterns with spoke-like structures in addition to the multiple-ring structure [79]. Recently, it has been reported that the DNA strand length (with the low and high number of base pairs) influences the drying dynamics, coffee-stain effect, formation of nanoscale structures, and aggregation. Interestingly, it also suggests a weak link between DNA viscosity and drying behavior [81]. Another study found that weak interaction forces between the synthesized hematite nanoparticle and DNA play an essential role in forming lyotropic liquid crystalline structures during drying [82]. The drying droplet method is also used to quantitatively measure trace DNA in a sample by tuning the substrate as a colloid-crystal surface and using a smartphone to develop a coffee-ring-based assay [58]. This study supports an earlier study where the authors qualitatively demonstrated the technique's sensitivity to trace amounts of DNA in the solution [79]. Recently, it has been reported that drying-induced evaporation can lead to different microstructural DNA arrays [14], which provides a platform to prepare and study different hierarchical structures. Figure 5(I-IV) presents a range of patterns formed due to drying of DNA and its mixture.

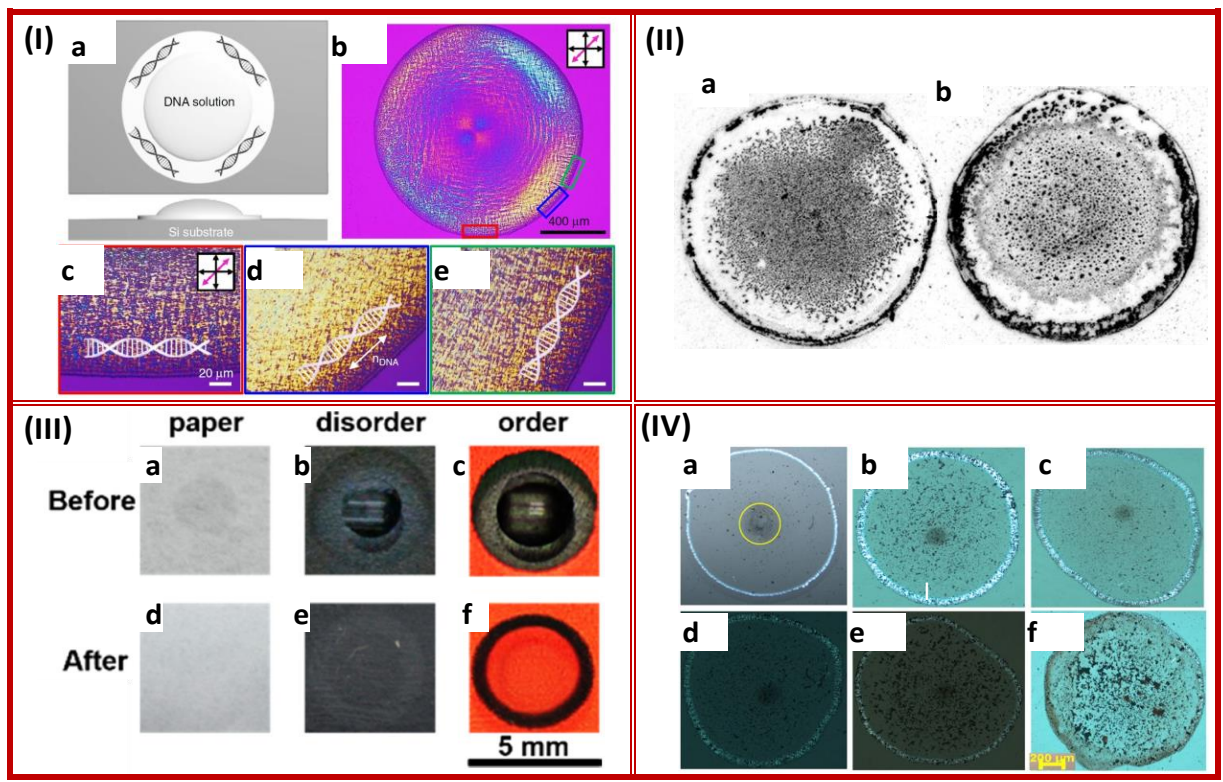

FIG. 5. (I) (a) Schematic of the droplet, (b) dried DNA droplet, (c–e) Zoomed images from b displaying different positions of the DNA droplet, and the corresponding orientation of DNA chains, adopted from [14]. (II) DNA dried droplets, containing 0.01% w/v with (a) 100 base pairs and (c) 1000 base pairs DNA chains, adopted from [81]. (III) DNA droplets after the LAMP reaction before (a-c) and after (d-f) drying on different substrates (paper and disordered and ordered SiO₂ colloid), adopted from [58]. (IV) (a) Dried droplet of DNA, (b–f) DNA with hematite NPs at different initial concentrations of 2.5, 5, 10, 20, and 50 mg mL⁻¹, respectively, adopted from [82].

RNA and other nucleic acids are expected to affect the drying pattern formation as well; however sessile configuration of such systems remains to be investigated. It will be insightful to compare the emerging patterns and link them to the native states of the RNA, DNA, and proteins. Studying the phase-separated droplets by RNA-binding proteins and RNA-DNA complexes is a significant missing link in the context of drying droplets, yet such experimental studies are rare.

## B.  Drying droplets of protein systems

### 1.  Drying droplets of globular proteins

Here, we will primarily discuss the experimental work on the drying droplets of different types of proteins. Surprisingly, the work on fibrous and composite proteins (the mixture of globular and fibrous proteins) in the sessile drying droplet configuration has not received sufficient attention in recent reviews. The work on the one-component, simple system, such as protein+DI, will be discussed first, followed by the multi-component systems, such as proteins+salts, and proteins+LCs (LCs are the phases whose properties lie between the solid crystalline and the (isotropic) liquid state [84]).

Gorr et al. [85] studied the evolution and the morphological patterns of Lys dissolved in water, varying the initial concentration from 0.1 (1 mg mL$^{-1}$) to 1 wt% (10 mg mL$^{-1}$). This study concluded that all droplet types exhibit a 'coffee-ring' effect [7]. The volume fraction of the Lys is found to be linearly dependent on the initial concentration; however, morphologically, the ring does not show any significant change in the ring's height and width. Later, the same group investigated the dependence of the pattern formation on the droplet size [86]. A simple cap-shaped deposit is observed for droplet diameters < 20 $\mu$m, whereas, ring-like deposits are found for diameters > 20 $\mu$m, depending on the initial Lys concentrations. However, higher-initial protein concentrations received limited attention. After some years, two fundamental questions emerged which aimed to explore (a) the role of the protein properties (in terms of mass, composition, configuration, and size); and (b) examine the effect of higher initial protein concentration (above 60 mg mL$^{-1}$ and up to 150 mg mL$^{-1}$) on the aggregation process. Figure 6(I) shows how the excessive aggregation of these proteins plays a role in relieving mechanical stresses during the drying process and determines the formation of the crack patterns. Interestingly, the uniqueness of the patterns in Lys+DI and BSA+DI is attributed to the nature of the protein particles, wherein the shape and size matter [33]. To understand the effect of the substrate temperature, the same group [32] has examined the tuning of the substrate temperature and initial Lys concentrations on the drying evolution and final morphology. Based on the droplet dynamics, three differ­ent concentrated regimes are found: ultra-concentrated, concentrated, and diluted. From a morphological point of view, the droplet forms the thickest film with suppressed coffee-ring formation in the ultra-concentrated regime at an elevated temperature. A slightly different protein+water system, compared to what had been done so far, was investigated where the authors directly used the egg-white solution [87]. The daisy and wavy-ring patterns are identified where the initial concentration could tune the nature of these cracks.

Building up the complexity of the components, researchers also investigated drying droplets that were prepared without de-ionized water (DI) to understand the ion-mediated effects on the pattern formation. Gorr et al. [88] demonstrated three zones (regions) from the periphery to the central regions in Lys+NaCl+DI systems: (a) a protein-rich amorphous peripheral ring, (b) a secondary ring-like area consisting of larger protein aggregates, and (c) dendritic and cubic crystallites. As NaCl concentration increases, the coffee-ring width was also found to go up. Similarly, in the BSA+NaCl+DI system, the lower initial concentration of BSA showed fine polycrystals, while upon increasing it, the salt crystallized primarily into various dendritic structures [89]. Not only that, HSA+NaCl+DI and HSA+DI systems were also compared: a chaotic crack pattern is obtained without the presence of any external salts, whereas the patterns are more ordered when saline water is used [90]. Based on a qualitative analysis, the authors argued that this uniqueness appeared due to the molecular features of HSA and the supramolecular organization of the polymer chains in a protein.

Interestingly, the mixture of Lys and BSA in DI is also investigated [91], where complex structures, such as small crystal aggregates, and dendritic patterns in the central region without adding external salts, are reported. This study also establishes how the structural changes in the morphological patterns can be used to detect the native (folded) and dena-tured (unfolded) states of these proteins. It leads to a question of whether the dendritic patterns emerge from the addition of external salts; possibly, they could be responsible for some folding mechanisms of these two globular proteins [91]. In a similar context, the same group studied the dried patterns to detect unfolded BSA, with the small amorphous aggregates forming at a low initial concentration of denatured BSA, compared to the eye-like pattern emerging at high concentrations [92]. The effects of using two different salts ($MgCl_2$ and KCl) in BSA+DI drying droplets has also been studied [93].

Researchers quantified these pattern-recognized structures in a similar complexity line using image processing techniques. For example, the pattern recognition algorithms are used in the Lys+NaCl+DI system to differentiate between deposits by extracting features using the Gabor wavelet and k-means clustering algorithms [94]. After training the images, the results show 90–97.5% classification accuracy in identifying these features [94]. The texture analysis [First-order Statistics (FOS) and Gray Level Co-occurrence Matrix (GLCM)] is also used to characterize the different stages during the drying process and complex deposited morphological patterns [see Figure 6(II)]. The FOS and GLCM parameters follow simple exponential laws which change as a function of the NaCl concentration [95].

In addition to the influence of these salts, drying patterns of proteins-NPs binary mixtures were also studied, though reports on these systems are scarce. The authors in [96] discuss the sensitivity of the drying method to detect a single-point mutation in a protein by forming unique patterns in native and mutant hemoglobin. In a similar context, the drying droplet method is exploited to predict the percentage of $\beta$-sheet content in HSA as a rapid quantification method. A distinctive pattern evolves with the increase in the $\beta$-sheets [97]. The birefringence and crystallinity of the emerging dendritic structures are discussed in the presence of magnetic NPs. The drying process influences the colloidal composites which self-assemble to form well-organized, multi-branched flower-like aggregate patterns [98].

In recent years, liquid crystals (LCs) have been used as a probe for protein solutions, wherein topological defects (regions where the average LC-molecular orientation, or the director, is locally undefined) have been observed. In BSA+LCs+DI droplets, a single defect is observed, while the LCs remain randomly oriented in Lys+LCs+DI. The presence of an umbilical defect of +1 strength in every domain near the edge of the BSA droplet is a surprising effect, which is mediated by the drying process [see Figure 6(III)]. The crack spacing in the dried Lys droplet is reduced in the presence of LCs, whereas no significant difference has been reported in dried BSA droplets [34]. The LC molecules are randomly orientated in the Lys+LCs+DI system, where each domain shows a central dark region surrounded by bright regions under crossed polarized conditions. The patterns get more ordered for the LC-added droplets compared to only Lys+DI [99]. Based on the results, it has been hypothesized that lightweight proteins give rise to randomly oriented LC director fields, whereas heavy proteins give rise to topological defects [100]. In the complex multi-component system, such as Lys+LCs in the presence of Phosphate Buffer Saline (PBS), examines whether the textural parameters can be correlated to the (a) phase separation of the salts present in the PBS and (b) LC birefringence during the drying evolution [35].

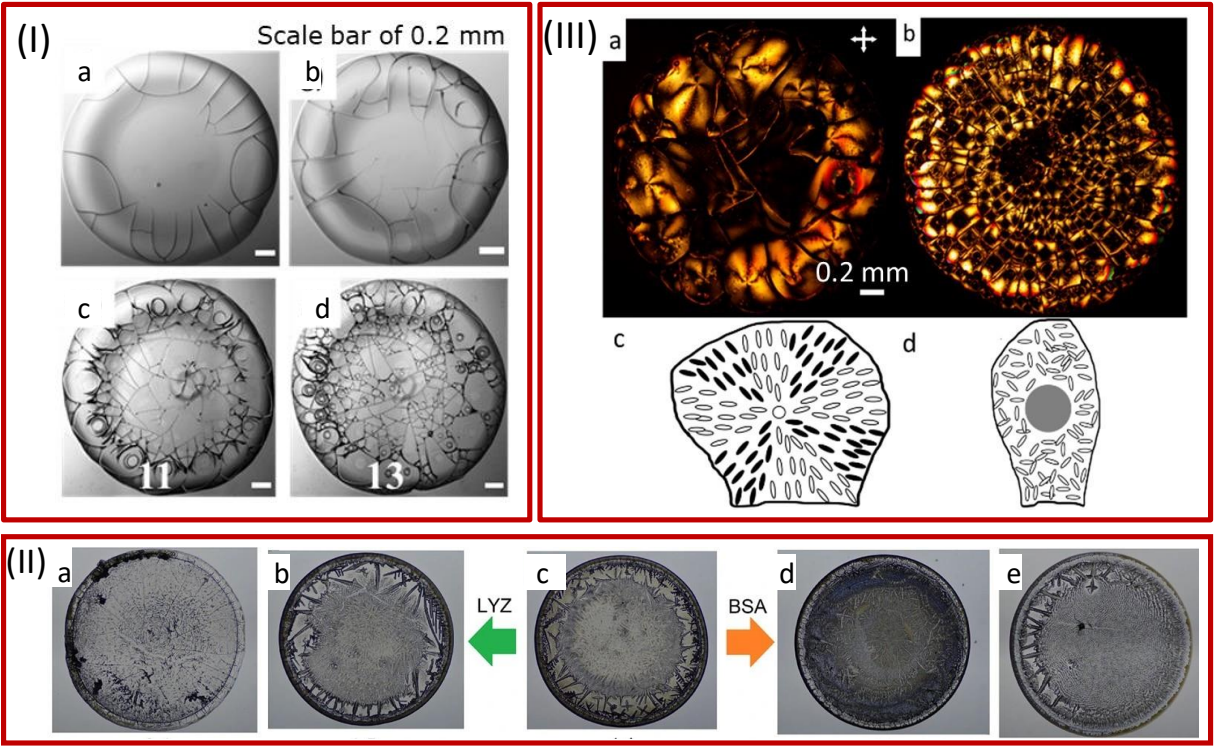

FIG. 6. Dried morphological patterns for globular proteins: (I) (a-b) BSA+DI and (c-d) Lys+DI at different initial concentrations of 11 and 13 wt%, respectively, adopted from [33]. (II) Umbilical defect formation in BSA+LCs+DI, (b) random distribution of LCs in BSA+LCs+DI, and (c-d) director configurations for (a-b), respectively, adopted from [34]. (III) BSA+Lys mixture at different initial relative concentrations of (a) 0:1, (b) 1:5, (c) 1:1, (d) 5:1, and (e) 1:0, adopted from [95].

*2. Drying droplets of fibrous proteins*

To the best of our knowledge, the first work on drying collagen droplets was reported in 1999. The authors showed a critical (minimum) collagen concentration for phase transition with and without patterns. Below this critical concentration, ordered and layered patterns were observed, and concentric ring patterns were captured above this concentration. This paper also discusses the birefringence properties and the onset of mechanical instability near the droplet periphery [101]. Another study investigates the influence of the drying rate and initial collagen concentration on pattern formation. It shows three distinct regions and patterns: anisotropic, a concentric ring, and a radially oriented region that the regulating parameters can tune. These patterns also form topographical clues inducing a cell (tenocyte) morphology, density, and proliferation [102]; see Figure 7(I). The effects of wettability and drying rate are reported in [103]. The increased hydrophilicity of the substrate promotes a reduction in the random distribution of the fibrils. A thicker layer of dried residue is observed for the hydrophobic case, while a lower deposition is found for the hydrophilic substrate. It also observed that the fibril length decreases with the increase in the drying rate. Researchers are also interested in understanding the evolution of drying patterns. It involves an oscillatory motion toward the central region (for the ring patterns) and aperiodic segmentation along the droplet periphery (for the radial patterns). The characteristic width of the radial bands exhibits a step-wise increase toward the pattern center [104].

Interestingly, the drying droplet method is also exploited for application purposes. For example, the Marangoni flows during the drying process direct the collagen fiber assembly that depends on the rate of self-assembly, RH, and droplet geometry. In response to aligned collagen networks, skeletal muscle cells promote orientation and differentiate into myotubes [105]. Another study shows how the drying droplet method can be implemented to fabricate periodic hierarchical structures of amyloid fibrils [106]. The periodic channels cause a stick-and-slip motion during the drying process. The development of the LC phase close to the pinned meniscus induces the fibrils to align while fiber-like structures are deposited uniaxially. The spacing between adjacent filaments, or their size, can be modulated by tuning the dimensions of the micro-templated substrate. Also, this method allows fabricating conductive microwires from AuNP-coated amyloid fibrils by exploiting the functionality of amyloid fibrils to nucleate inorganic NPs on their surface [106]. Recently, pattern-recognized

Machine Learning Algorithms (MLAs) have been examined to identify the features in the drying droplets [see Figure 7(II)]. The machine-based algorithm helps identifying the primary and secondary protein structures where the patterns result in the stratification of eight amyloid-beta (A$\beta$) variants with an accuracy > 99 %. Also, the patterns of a range of distinct A$\beta$-42 peptide conformations are identified with accuracy of > 99 % [107].

3. *Drying droplets of composite (mixed globular and fibrous) proteins*

We will now shift our focus to other types of proteins that cannot be classified as globular or fibrous proteins. The proteins present in dairy products (whey proteins and the casein micelles) and microtubules. It is worth mentioning that dairy proteins are extensively used in the spray drying method [108]. However, studies of the dairy proteins used in the spray drying method are beyond the scope of this review. The initial concentration influences the drying dynamics in whey proteins, typically showing three regimes: (a) sol-gel transition, (b) buckling of the droplet, and the formation of a protein shell, leading to the (c) formation of a vacuole [64]. A separate study investigating the drying dynamics of caseins+DI shows three stages of the drying process: (a) formation of a thin layer that undergoes (b) surface instabilities (buckling and invagination), and (c) formation of a deflated and wrinkled particle shape [109]. The same group extended the investigation by comparing the drying dynamics of whey protein and casein micelle droplets, reporting that different time and length scales underpin the observed mechanical instabilities (buckling and fracture). The protein's native characteristics explain the difference in their mechanical properties, i.e., casein micelles are soft and deformed colloids, whereas the whey proteins are hard spheres [63]. Another study examines the influence of the drying rate, suggesting that the skin evolution and its mechanical properties correlate with the drying rate; and that it also depends on the composition of the mixture (size, shape, and mass). Interestingly, despite different process characteristics, time scales, and experimental drying methods, the drying dynamics are similar [110]. Comparing the mechanical behavior through the crack formation and indentation testing of drying dairy and model colloidal systems confirms that the model system can be a powerful tool for revealing signature patterns for differences at the molecular scales [111].

Furthermore, the mixture of casein and whey protein drying droplets was investigated recently. Figure 7(III) exhibits the patterns of individual casein and whey protein dried

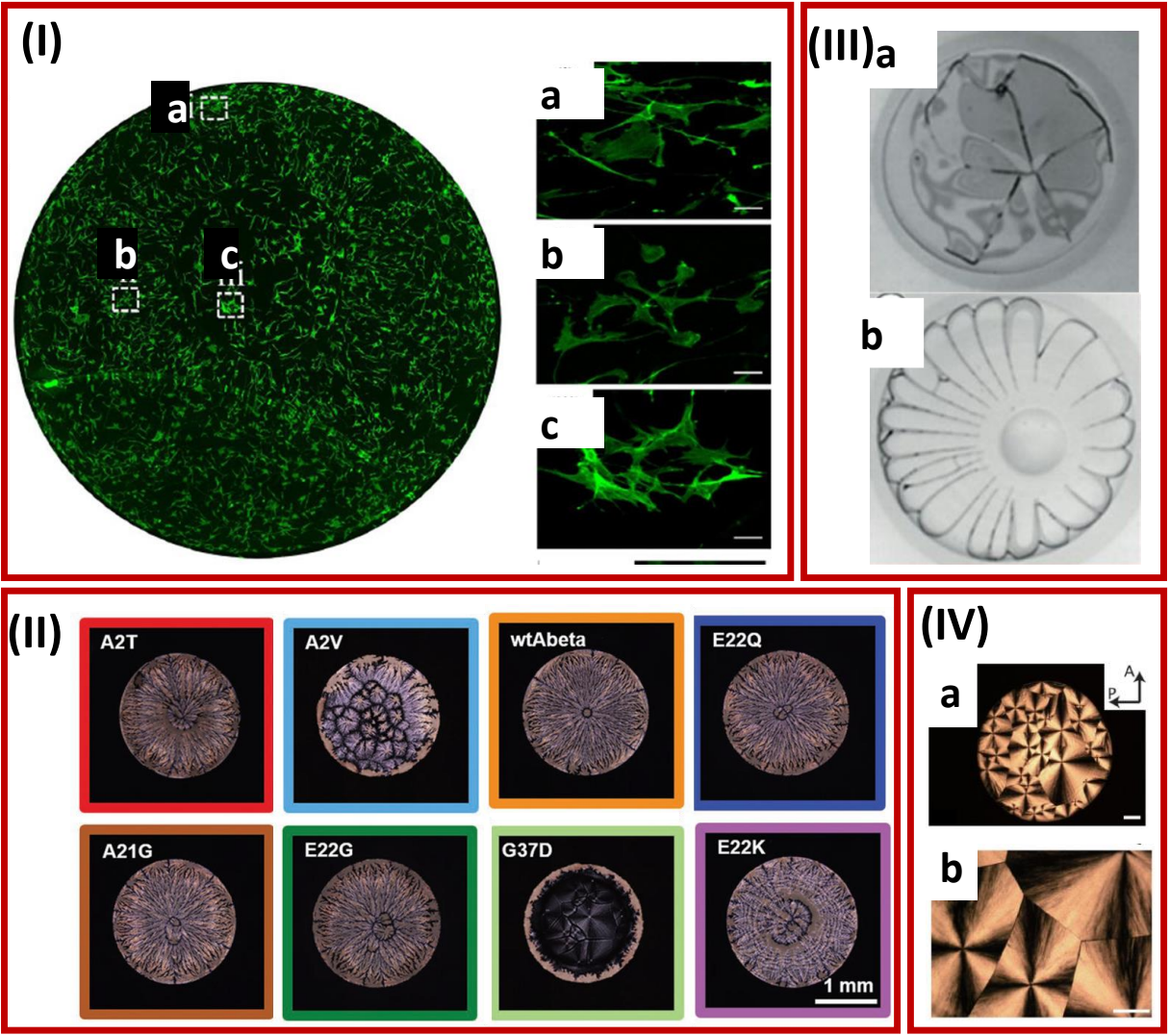

FIG. 7. Dried morphological patterns for proteins: (I) Collagen imaged at (a-c) different locations, adopted from [102], (II) A$\beta$42 peptide and its seven different variants with a single point mutation, adopted from [107], (III) (a) casein micelles (observed in the form of native phosphocaseinates), and (b) whey proteins (observed in the form of isolates), adopted from [111], and (IV) Microtubules under (a) 2×, (b) 4×, adopted from [65],

droplets in the form of isolates. The drying evolution and analysis of surface mechanical properties as a function of the whey and casein protein composition were investigated. The interfacial rheology using the oscillatory drop method confirms the protein segregation in the skin related to the bi-disperse colloid dispersions [112]. Another study includes a preliminary phase diagram by which the tuning and controlling of the initial concentration are possible. It predicts that this drying droplet method can be used as a simple, cost-effective laboratory scale approach to understanding milk abnormalities and its supplements [62].

Recently, a study on microtubule+DI droplets exhibits the interplay between initial concentration and substrate temperature, which is crucial for deciding the drying-induced morphological patterns. The spherulite form [see Figure 7(IV)] is found in the highly concentrated tubulin solution via nucleation and subsequent polymerization. These microtubules are also aligned radially and can be tuned by temperature-induced polymerization and depolymerization. Interestingly, this structure is also optically birefringent [65].

### C. Bio-fluids: Recent advances in drying droplets

*1. Drying droplets of blood, blood plasma, and serum*

We will describe the recent advances in blood plasma, serum, and blood over the last five years. Older research on this topic is well-documented in previous research articles [41, 113]. Here, we will focus on summarizing the fundamental studies which were recently published. Many fundamental experimental studies have been carried out on drying blood droplets [113–119]. Recent works have focussed on (a) phase separation when the serum spreads outside the main blood pool and (b) the competition between coagulation and drying at different temperatures and RH ranges. These works establish that the increase in RH enlarges the final pool area. The serum spread outside the main pool at RH > 50% at elevated temperatures. Interestingly, this phase separation is more significant on varnished wooden surfaces and does not lead to any changes in the drying morphology [120]. The key feature of these varnished wooden floors is that these surfaces are made smooth by applying three layers of varnish. Since it is not yet possible to make bio-mimetic blood like artificial urine [121], an attempt is made to compare the RBCs in PBS, BSA in PBS, and polystyrene colloidal particle model systems to identify the vital mechanistic parameters for explaining the resulting morphological patterns. The high RBC concentration does not suppress the coffee-ring effect as the high polystyrene colloidal particles do [122].

Surprisingly, investigations on the dilution of human whole blood are largely missing. Concentration-driven phase transition has been reported during the drying evolution and in dried droplets when blood is diluted from 100% (undiluted whole blood) to 12.5% (diluted concentrations). It is due to the visco-elasticity that the dilution decreases, ultimately dis-appearing at the dilution concentration for the observed phase transition equivalent to 62% (v/v) [36]. The same group has extended this study to protein and blood systems. The effect of different substrate temperatures is also examined on the pattern formation. The transition behavior is not limited to multi-component bio-colloid (blood) only but may be a phenomenon of a bio-colloidal solution containing many interacting components. Interestingly, the high dilution of blood behaves like the BSA solution. The higher drying rate due to elevated temperature shows that the components do not get enough time to segregate and deposit to the substrate, unlike ambient conditions [31]. The substrate's mechanical properties, pattern formation, and wettability are related to the diluted blood for the first time. It exhibits that blood droplets drying on a hydrophilic substrate can give rise to radial and orthoradial cracks in the coronal region and random cracks in the central region. Buckling instability is observed for the hydrophobic substrate, which can be attributed to the distinct wetting and drying kinetics as the diluted blood droplets get dried under uniform conditions [123].

In the context of plasma drying droplets, the exploration is done on the plasma and plasma+polymer mixtures. It is found that the internal flow of a drying plasma droplet is not the same as that of the water droplet in both sessile and pendant configurations. For water, only capillary flow exits, whereas the capillary flow and the buoyancy-driven natural convection co-exist. It is argued that this phenomenon occurs possibly due to the adsorption of the plasma proteins onto the plasma-air interface, eliminating the surface tension gradient and the Marangoni flow [124]. The plasma and plasma+polymer system inspection show the spectral differences between the inner and outer-ring regions in the dried samples. Interestingly, patients with different cardiac-related diseases are also investigated. It reveals that the distribution of biomolecules is distinct in each region of the droplet, and the unique pattern is correlated with the patient sample [125]. The effects of temperature and the RH are also investigated on the drying patterns of blood serum and plasma extracted from different human samples. It is found that the lower RH triggers crack initiation and formation. In contrast, higher RH increases the crystalline structures, mainly in the central region of the droplet. Interestingly, these structures correlate with different contents and compositions of the blood [126].

Beyond human samples, recently, researchers have used fetal bovine serum [127, 128] to understand the pattern formation during drying. It describes different zones in the mor-phological patterns: a peripheral protein ring, a zone of protein structures, a gel, and a central crystalline zone. The area of the crystalline zone is found to shrink with the increasing substrate temperature. These structures change from dendritic to cell-like at elevated temperatures. Furthermore, the drying evolution, final deposit volume, and cracking are connected. It is found that the coffee-ring volume decreases with this increasing temperature, whereas the number of cracks increases due to the faster drying rate at elevated temperatures.

Recent progress is also made in blood serum drying droplets where the NPs are added as a probe in the system. Using a multi-disciplinary approach that combines (a) the use of MLAs and (b) microscopy and Surface-enhanced Raman scattering (SERS), gold nanoparticle colloid mixed-blood serum samples from patients with colorectal cancer were investigated. SERS spectra analyzed the resulting coffee-rings. Accurate MLAs were crucial for the detection of cancerous serum [129]. In a similar context, droplets of silver NPs-serum mixture by SERS and droplet drying method in prostate cancer patients were also studied. The droplets from the coffee ring: the ring is suppressed in control cases, enhancing the stability and reliability of SERS detection. Different MLAs, such as Partial least square (PLS) and SVM algorithms, yielded a high accuracy of 98.04% to distinguish the typical sample from the two types of cancers [130].

## 2. Drying of bio-mimetic and real salivary droplets

Triggered by the recent COVID-19 pandemic, there is also a growing interest in the colloidal community to study the pattern formation driven by the drying process. Interestingly, bio-mimetic fluids of human saliva are used in most of the studies. For instance, researchers have investigated the drying dynamics, precipitation, and the resulting pattern formation in the bio-mimetic fluids influenced by different RH levels. However, it is to be noted that the pH and texture of the bio-mimetic fluid (NaCl, mucin, with and without DPPC) do not match with the human saliva. The fluids are opaque with a pH of 3.7, whereas the saliva is clear with a pH of 7.5 [131]. Another study uses a similar bio-mimetic fluid, but researchers are interested in understanding the fluid's pattern formation on different substrates (glass, ceramic, steel, and plastic). The virus-emulating particles (VEP) of nanometer size mimic the virus. The dendritic and cruciform-shaped crystal patterns are found in all substrates except steel, where the regular cubical crystals are formed [132]. In a similar context, 100 nm latex NPs are used to mimic the virus, and studied the desiccated, precipitated, and crystallized surrogate droplets in relation to the unique pattern formation [133]. However, to get the actual essence of what happens, we need to use natural human saliva in the presence of bacteriophage. Along this line, a study is published where the virus viability assays in human saliva are measured to examine the survival of three bacteriophages (the enveloped Phi6, non-enveloped PhiX174, and MS2) at various RH (23, 43, 57, and 78%). It is concluded that only the RH and droplet hydration states are insufficient parameters to explain the virus survival. The virus-suspended medium and the interaction between virus-saliva components play a significant role in the virus survival [134]. Furthermore, the evaporative-drying dynamics of the bio-mimetic fluid are studied by semi-levitating the droplet on the superhydrophobic substrates. The droplets have a minimal solid-liquid contact area in this configuration. This study systematically compares the drying dynamics and the patterns for pure water, salt solution, salt water with mucin, and surfactant-added solutions at different RH. These droplets' stability is compared to understand the generating conditions for the virus survival [135].

## IV.   FROM DRYING PATTERNS TO CLINICAL DISORDERS AND DISEASES

As discussed earlier, blood is compositionally complex and heterogeneous due to the presence of cellular components (RBCs, WBCs, and platelets) and their self-assembling mechanisms. Any change to the native states of the individual components or the physical and chemical changes leading to their shape deformations during evaporation might alter the heterogeneity of the blood. Therefore, distinct patterns emerge for the different physiological conditions.

For example, the physical property of the blood (viscosity) decreases,  and the blood gets thinner than the healthy blood [136] when a patient suffers from very low platelet counts, often seen in dengue or malaria patients [137]. Whereas, in the case of sickle cell anemia, increased RBC aggregation is associated [138]. These patterns have been used for Blood Pattern Analysis (BPA) in forensic science, where the patterns were investigated for identifying crime details [139]. Though the underlying physics in the drying droplets of blood gained colossal popularity from Brutin's work in 2010, interestingly, H. L. Bolen reviewed the investigations done from 1939 to 1953 and suggested that a distinct pattern for each disease could be a potential marker for diagnosis of any malignant diseases [140]. Archeological analysis has also been performed in many cases to understand the cellular components of the blood from ancient times [141]. Thanks to the advent of MLAs to identify those patterns, there is a growing interest in linking these patterns for bio-medical diagnostic purposes. However, studies have seldom been conducted to understand the causes of specific patterns due to a particular disease. One can only speculate such an evolution of pattern is linked to changes in the composition of biomolecules, their self-assemblies, and other solutes. The field is still growing, and here we summarize the drying droplet investigations related to pattern-recognized diagnosis. In this section, for the first time, we will provide a catalog of blood serum, plasma, and blood droplets elucidating how the dried morphological patterns appear for (a) different abnormalities and disorders and (b) diseases over their different stages.

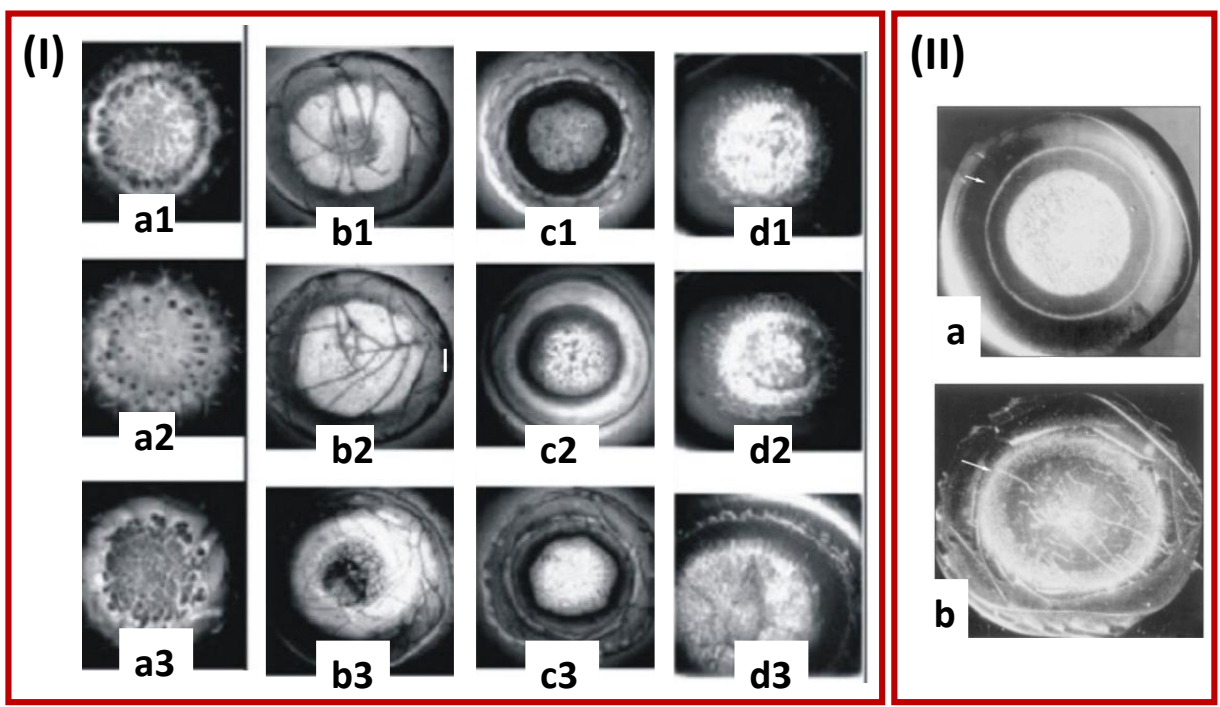

FIG. 8. Dried Patterns of blood serum and plasma droplets taken from women before, during, and afterthe pregnancy period: (I) Blood serum droplets for (a1-a3) healthy, (b1-b3) in-time de-livery, (c1-c3) premature delivery, and (d1-d4) premature delivery in different periods of gestation, adopted from [142]. (II) Blood plasma in the early postpartum period: (a) normal childbirth (after 40 weeks of gestation), (b) premature delivery (childbirth after 34 weeks), adopted from [143].

Researchers have exploited the pattern-recognized drying droplet to find a relation between the bio-fluid's abnormalities, patterns, and physical characteristics. Figure 8(I-II) shows that these patterns differ for pregnancy periods. A new technique based on Acoustic Mechanical Impedance (AMI) of a droplet drying on a solid substrate of a quartz resonator oscillating with ultrasound frequency is established, capturing the drying dynamics [142, 143]. The AMI curves demonstrate that the structuring and the self-assembly (or aggregation) of the plasma droplets are influenced by normal pregnancy vs. when premature birth or threatened premature delivery is diagnosed. However, this study fails to provide any concrete underlying mechanism. It relates these differences due to the abnormalities of the pregnancy-associated proteins (macroglobulins), hormones in the plasma, and the viscosity of the blood.

A schematic diagram is presented in Figure 9. The diseases are described in the increasing levels of complexity, such as from bacterial-viral infection to diabetes and related morbidity issues, and thalassemia to cancer [see Figures 10-14].

Recently, a landmark paper is published in the context of pattern-recognized disease pre-diagnosis [144], establishing a link between the physio-chemical properties of the bio-fluid and the unique pattern formation. The blood-drying droplets of healthy adults and infants with jaundice were investigated. It is found that the crack length and height profile of the dried droplets are different for these groups. More so, a relation is established between the crack length of the jaundice group and the blood's bilirubin levels [see Figure 10(I)]. In contrast, if we compare Figure 10 (I) and (II), surprisingly, these patterns are not the same. However, both the anemic blood patterns are found to be different from the healthy ones. It might be because of the different substrate (glass vs. aluminum tiles) and environmental conditions (T = 22˙C vs., T = 32˙C) [66, 136].

The blood serum [see Figure 11(I-II)], plasma [see Figure 11(III)], and blood droplets [Figure 11(IV-V)] are deposited on the substrate. The patterns are found to be distinct for the bio-fluid extracted from healthy vs. viral and bacterial-infected patients [136, 142, 147, 148]. Recently, the standard smear pathological methods used in diagnosing TB have been compared to the Marangoni-based flow patterns in a drying droplet configuration [149]. It is to be noted that different types of bio-fluids matter cannot compare the patterns that emerged in the drying droplets of serum, plasma, and blood. Of course, the highest reproducibility is found within the study; however, we cannot compare the patterns for different types of hepatitis and Tuberculosis. Furthermore, we do not have the harmonized protocol; for example, Figure 11(IV-V) shows the patterns for the blood droplets dried at (substrate = aluminum, T = 32°C and RH = 50% [136]) and (glass slides, T = 20 − 22°C, and RH = 63-71% [148]), respectively.

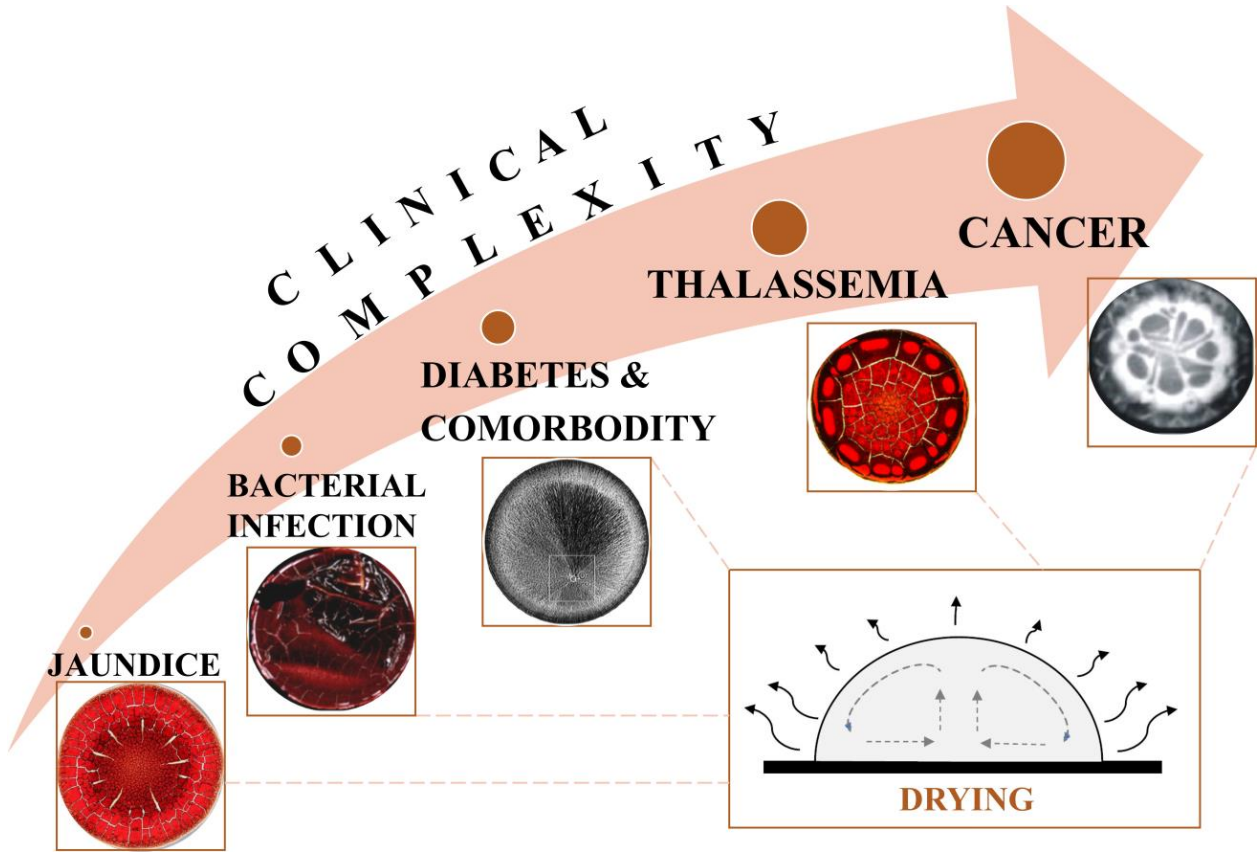

FIG. 9. A schematic representing the dried patterns of bio-fluids from different diseases with increasing complexity. The images for jaundice, bacterial infection (TB), diabetes, thalassemia, and cancer are adopted from [144], [136], [145], [146], and [142], respectively.

Figure 12(I-III) exhibits the droplet dried patterns in diabetic and comobormodity issues for different bio-fluids (lysed blood [145], vs. blood serum [142], vs. blood [66]. Since the pattern of diabetes has emerged from drying droplets of lysed blood added into the cupric chloride aqueous solution, it is hard to conclude whether this is a fingerprint of any diabetic samples. The related abnormalities often found in diabetic patients are paraproteinemia and hyperlipidemia. Paraproteinemia indicates the presence of excessive amounts of immunoglobulin proteins in the plasma that causes damage to the kidneys [Figure 12(II)]. Hyperlipidemia indicates the presence of high cholesterol due to an excess of lipids or fats in

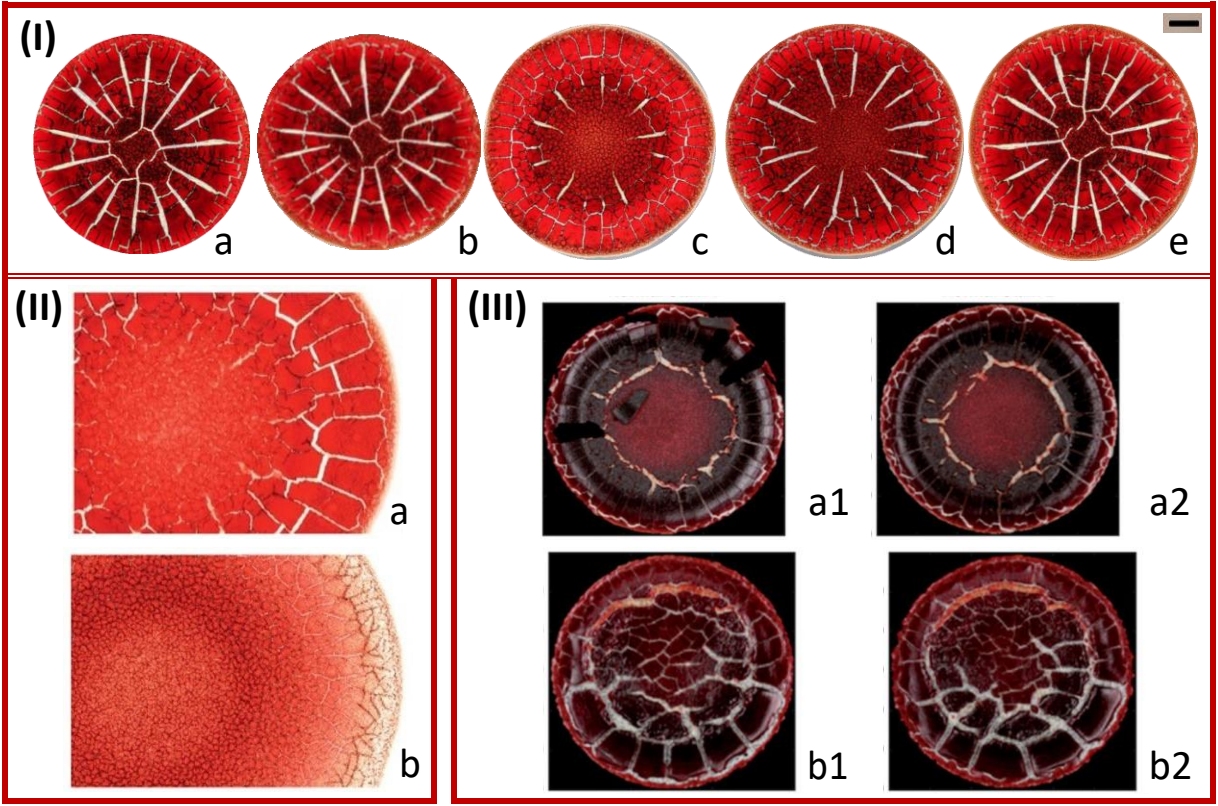

FIG. 10. Patterns of whole dried blood with diseases affecting babies and young adults: **(I)** (a) healthy young male, (b) a male infant with jaundice, babies with jaundice at different concentrations ($\varphi$) of total bilirubin (c) baby boy with $\varphi$ = 7.1 mg dL$^{-1}$, (d) baby girl with $\varphi$ = 13.4 mg dL$^{-1}$ and (e) baby boy with $\varphi$ = 29.3 mg dL$^{-1}$, adopted from [144], **(II)** (a) sample from a healthy young woman, and (b) Woman with anemia, adopted from [66], and **(III)** (a1-a2) healthy, and (b1-b2) anemic, adopted from [136].

the blood [Figure 12(III)]]. These images, no doubt, establish the uniqueness of the emerging patterns in the drying droplets; however, the studies are in the nascent stage.

A fundamental study on understanding the distinct pattern formation in the dried blood droplets taken from thalassemia patients [Figure 13(I-II)] is presented. Figure 13(II) from [144] depicts that similar crack patterns are formed in these samples, establishing a quantitative feature (crack length) for distinguishing the patterns between healthy and patients. Four years later, a similar study was done, where the authors showed how the underlying complex mechanism of cell-cell and cell-substrate interactions play a dominant role in deciding these patterns [146]. However, it is to be noted that the texture of these thalassemia and healthy samples are not exactly the same [Figure 13(I-II)]. It might be that age is critical, or even gender, as a diverse set of individuals are involved. Since there is a vast source of variability, one needs to perform the experiments systematically and carry out rigorous statistical tests to determine the significant features (age, gender, geography, ethnicity, etc.) in pattern formation. This is an open question, and currently, we lack critical evidence to prove if and how these parameters (sources) contribute to the unique patterns.

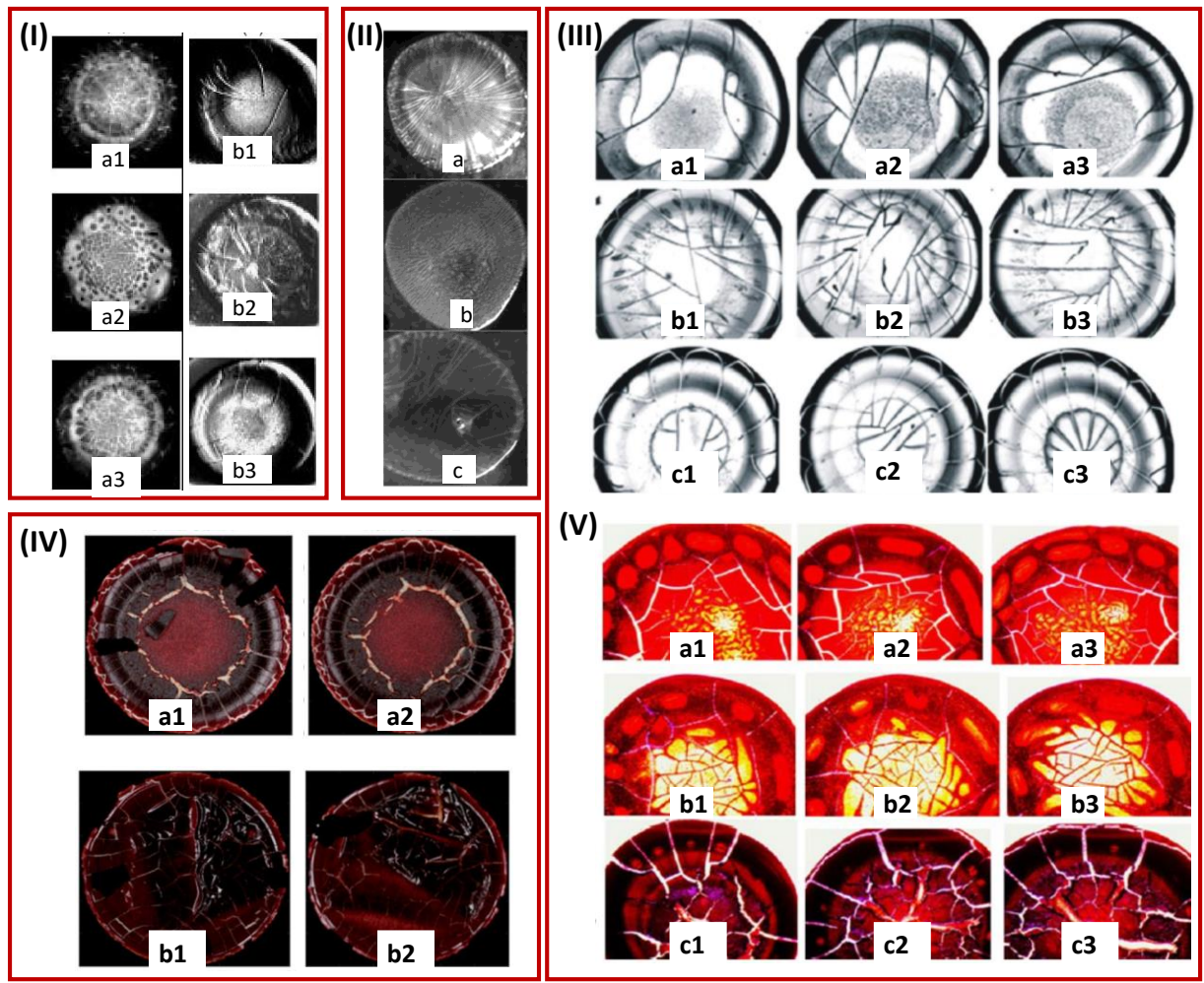

FIG. 11. Dried Patterns of different bio-fluid droplets for the diseases infected from bacteria and viruses: Blood serum dried droplets for (I) (a1-a3) healthy, and (b1-b3) hepatitis, adopted from [142]. (II) (a) healthy, (b) hepatitis B, and (c) hepatitis C, adopted from [147]. (III) Patterns in dried blood plasma droplets for (a1-a3) healthy, (b1-b3) bovine leucosis virus positive, and (c1-c3) bovine tuberculin positive adopted from [148]. Patterns formed in blood-dried droplets in (IV) (a1-a2) healthy, and (b1-b2) TB-infected patients, adopted from [136], and (V) (a1-a3) healthy, (b1-b3) bovine leucosis virus positive, and (c1-c3) bovine tuberculin positive adopted from [148].

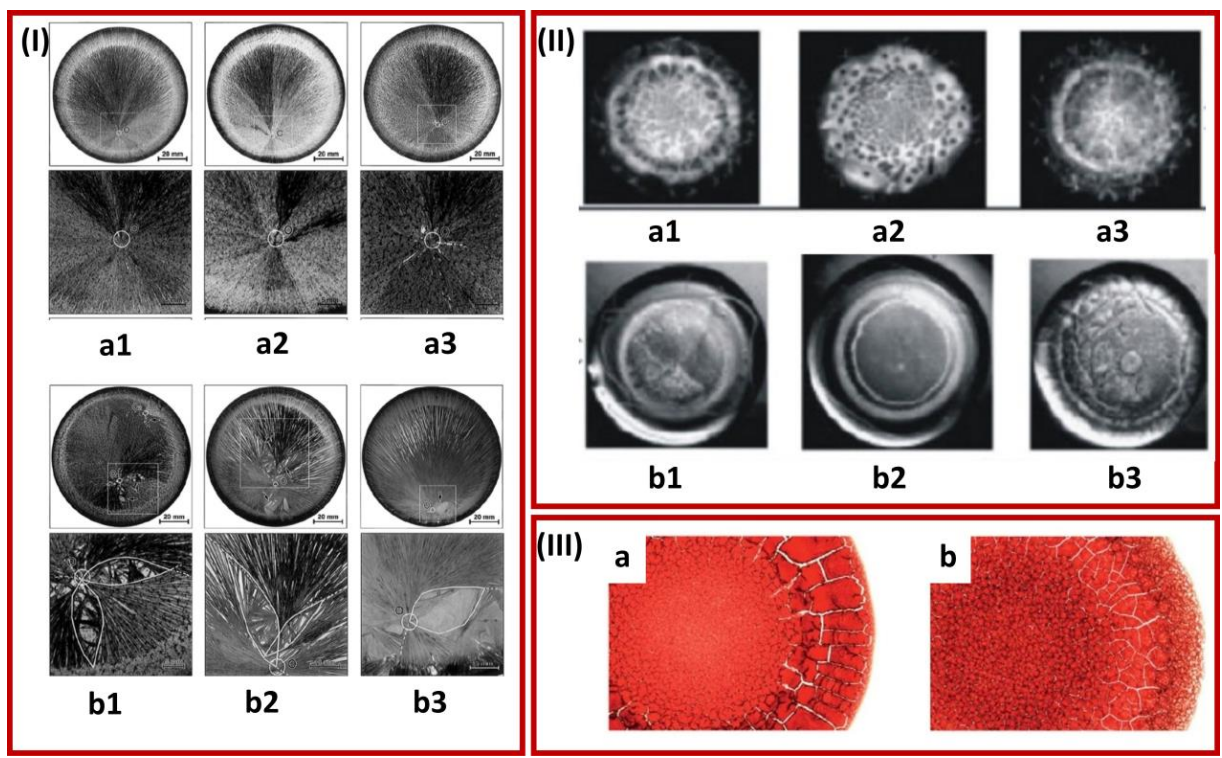

FIG. 12. Droplet dried pattern formation in diabetic and related disorders and diseases for different bio-fluids: (I) Patterns of dendritic crystal growth in cupric chloride aqueous solutions to which a slight amount of lysed blood from diabetic patients (a1-a3) and healthy persons (b1-b3), as controls, were added *in-vitro*, adapted from [145]. (II) Dried blood serum droplets for (a1-a3) healthy, and (b1-b3) paraproteinemia, adopted from [142]. (III) Dried blood droplets for (a) healthy young man and (b) young man with hyperlipidemia, adopted from [66].

Reproducible morphological patterns of blood serum droplets for healthy people and different types of cancers, including breast cancer [142], lung cancer [142], and carcinoma [151] Figure 14(I, III), have been reported. Furthermore, the different dosage of chemotherapy is also found to impact the emerging patterns in plasma-dried droplets [150]; see Figure 14(II). Though these images show reproducibility, no quantitative comparisons or rigorous statistical tests were performed. The cracks and texture are different for different cases, so one can do further image analysis to quantify the different features. However, high-resolution images are necessary for quantitative image analysis. Furthermore, the drying substrate, along with the drying kinetics, significantly affects the pattern formation rather than the composition itself of the bio-fluids.

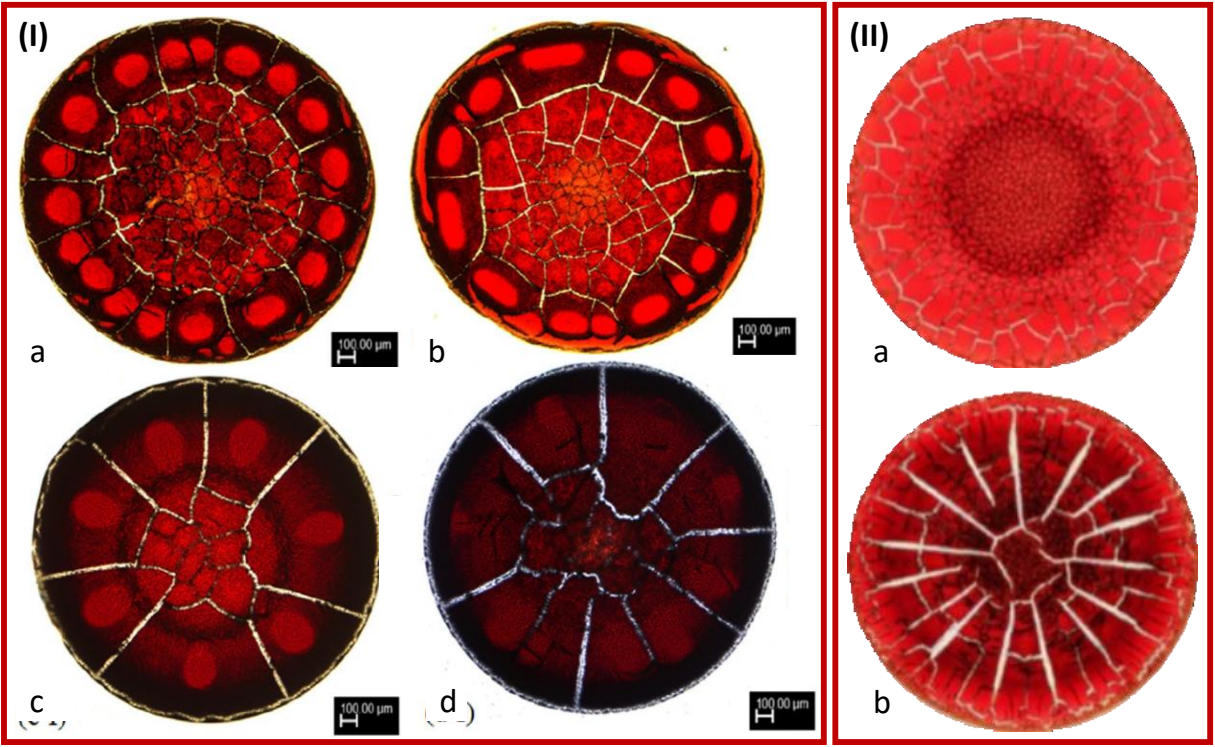

FIG. 13. (I) Patterns of whole blood dried droplets of: (a-b) thalassemia and (c-d) healthy samples, adopted from [146]. The samples are taken from two different individuals, (II) (a) a 19-year-old male with thalassemia, and (b) a healthy young male, adopted from [144].

Table III presents a comparative summary of the type of disease, disorder, and abnormality. The table also shows the substrate on which the droplet is deposited and lists the environment during the drying process. The established (qualitative and quantitative) pattern recognition techniques and how these patterns are related to the physical characteristics of bio-fluids (if any) are also shown in this table. The strengths or weakness is also summarized here.

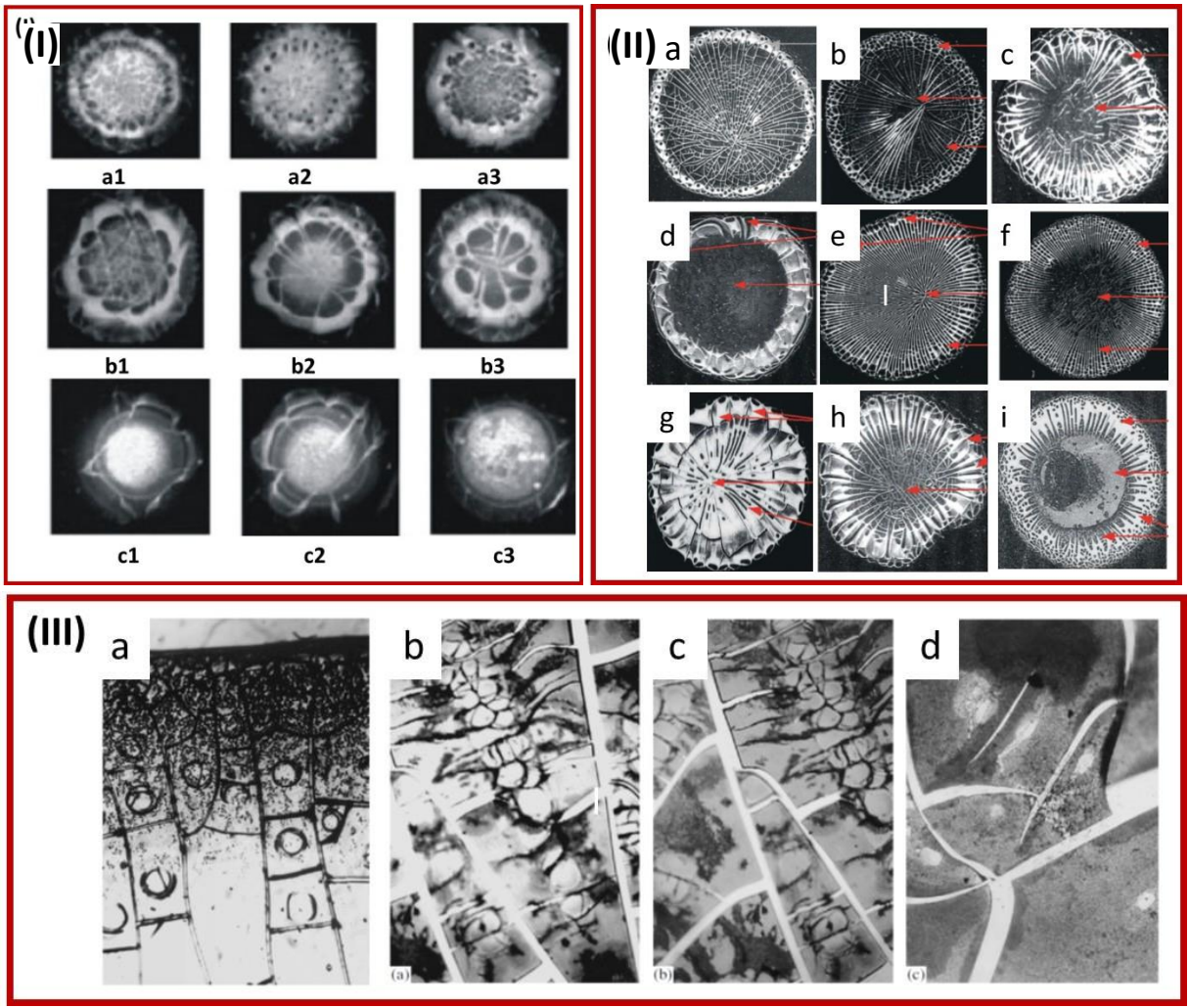

FIG. 14. Dried Patterns of blood serum and plasma droplets for cancer: (I) Blood serum dried droplets for (a1-a3) healthy, (b1-b3) breast cancer, and (c1-c3) lung cancer, are shown here. This figure is adopted from [142]. Dried droplet patterns of blood plasma from (II a) healthy, (b) 1st morphotype of ILF, (c) 2nd morphotype of ILF, (d) 3rd morphotype of ILF (e) 4th morphotype of ILF, (f) 1st morphotype of IIP, (g) 1st subtype of 3rd morphotype of IIP, (h) 2nd subtype of 3rd morphotype of IIP, (i) 4th morphotype of IIP, adopted from [150], and (III) (a) healthy, and (b-d) carcinoma (cancer that forms in epithelial tissues), imaged at different sections of the dried droplet, adopted from [151].

TABLE III: A comparison of studies in the context of fundamental understanding, quantification, pre-diagnosis (diseases, and abnormalities) and the pattern formation in the drying droplets of different bio-fluids.

| Disease/ Abnormality | Substrate | Env. | Qualitative Analysis | Quantitative Analysis | Characteristics | Weakness or Strength | Refs. |
|---|---|---|---|---|---|---|---|
| Anemia and Tuberculosis (TB) | Aluminum tiles polished by sand paper | T = 32˙C and RH = 50% | Dried droplet thickness: It varies from the central (thin) to the periphery (thick) for healthy, whereas infected ones have a constant thickness, thus failing to distinguish between Anaemia and TB | Texture difference between the images of a dried droplet for a healthy person and patient | A plausible reason stated is the change in the mechanical fluidic properties of the blood (surface tension, wettability, viscosity and viscoelasticity) | 28 samples, a bigger set of specimens is needed | [136] |
| Anemia and Hyperlipidaemia | Glass slides | T = 22˙C | Anemia: near the edge, the images look light-colored compared to the central region. Hyperlipidaemia: near the edge, the images look thick and greasy. | | A plausible reason stated is that the dried pattern of an anaemic patient corresponds to a non-wetting situation of the blood | Quantification is needed | [66] |
| Thalas-saemia and Jaundice | Micro-scope cover glass | T = 22·C and RH = 45% | Healthy and jaun-dice group: Two thick rings near the contact line, Thalassaemia group: One ring and its thickness are low com-pared to the other groups. | Crack length | HCT, MCV, bilirubin con-centration and the plasma viscosity | A land-mark paper relating bio-colloids and pattern recognition | [144] |
| Thala-ssemia | Glass slides | T = 22·C | Central region of infected samples are dis-tinctly larger with more branching cracks, as com-pared to that of the healthy ones. | Crack length | HCT, MCV, and the plasma viscosity re-lating to cell-cell and cell-substrate interactions | A land-mark paper relating bio-colloids and pattern recognition | [146] |
| Different periods of pregnancy: normal pregnancy vs. premature birth or threatened premature delivery | Pure slides | T = 18 − 22˚C, and RH = 60–70% | | Plasma structuring index during the drying process | A plausible reason stated is that the drying pattern is associated with an increased content of pregnancy-related proteins (macroglobulins) in the plasma and viscosity | The image quality of the dried samples are poor, and no qualitative investigation is done | [142] |
| Viral infection | Glass slides | T = 20 − 22˚C, and RH= 63-71% | | Phase transitions during the drying process | | Quantification is not well-settled | [148] |
| Diabetes | Petri dish | T = 28˚C, and RH= 45% | Diabetic samples: Larger convex lens bodies near the nucleation center of the dendrites. Healthy samples: Small convex lens in size | Long diameter of the convex lens shape | | Lysed blood is added into the cupric chloride aqueous solution, a study is not performed on whole blood | [145] |

| Different types of Cancer | Pure slides | T = 18 – 22°C, and RH= 60–70% | | Differences in the dynamics of phase transitions during the drying process | A plausible reason stated is that the drying pattern is associated with an increased viscosity due to the increased content of monoclone immunoglobulins. | The image quality of the dried samples are poor, and no qualitative investigation is done | [142] |
|---|---|---|---|---|---|---|---|

Though these patterns show reproducibility within a study, discrepancies are observed when compared across different studies done by researchers worldwide. For example, if we compare the patterns of the whole blood extracted from healthy humans, it is observed that the texture and the length of the crack patterns are not the same [see Figures 12(III) and 13(I-II), and Table III]. This difference might be due to the initial conditions, i.e., whether there is any anticoagulant present in the blood, how the blood is stored, or any further processing involved before the deposition of these droplets on the substrate. In addition, substrate, drying, environment (temperature and RH) conditions, etc., also play an important role in deciding these patterns. Therefore, one needs to develop and put in place standardized protocols for sample extraction and preparations, controlling environmental factors and specifying image acquisition parameters for reliably using drying-based techniques in clinical settings.

## V.  DRYING OF MICROBIAL DROPLETS

Microbial active matter [20], including bacteria, algae, fungae, and other microbial community members like the bacteriophage and viruses, are ubiquitous in a range of natural and built-in settings, including hospitals, public lavatories, and residential settings. Microbial populations exchange energy and matter with their surrounding environment and, thus, are out-of-thermodynamic equilibrium. They fall under two categories depending on their ability to navigate actively: motile and non-motile species. Motility confers microbes an essential trait to migrate actively in response to external cues. Therefore, it will be consequential for dynamic environments like a drying droplet. Microbial droplets are highly relevant in our day-to-day lives and can often be observed in association with fomites. In this section, we will discuss the patterns that emerge due to *active microbial droplets*. Figures 15-19 display different patterns formed in the microbial droplets drying under various conditions.

To the best of our knowledge, one of the earliest works on drying microbe-laden droplets was carried out by Fang et al. [152], with the key idea of measuring bacterial survival on teflon-coated slides under different saline conditions. A year later, a follow-up study [153] analyzed bacterial suspensions of different strains to understand whether drying-induced pattern formation depends on bacterial motility. Microbial motility was identified as an essential parameter that mediates the morphology of the emerging patterns. The non-motile species lead to circular ring patterns near the periphery (with very few bacteria present in the interior region), and the motile species produce disc-like patterns. By increasing the motility systematically, the uniformity of the deposited disc-like patterns could be controlled. However, Sommer and Zhu [154] argued that bacterial deposition in the interior region (for the motile case) could arise due to (a) the differences in adhesion properties of the motile bacteria (motile bacteria is *stickier* than non-motile ones and (b) chemotaxis (the movement of the bacteria due to the concentration gradient of chemicals) promotes the bacteria towards the drop edge. The authors of [153] provided counterarguments that mutant strains were used for motile and non-motile bacteria, such that stickiness remained similar. The sterile washing protocol left the bacteria with no chemical gradients for chemotaxis [153]. Later, experiments by Kasyap et al. [18] supported the observations made by Nellimootil et al. [153], revealing that the pattern formation is related indeed to the motility of the bacteria and not influenced by chemotaxis and/or production of sticky bio-surfactants. In addition, a periodic variation in bacteria density (concentration) is noticed near the periphery. The authors argued that it might be due to the balance between evaporative convective flow and the diffusive dynamics of bacteria in the drying droplet. This concentration variation drives the fluid velocity and shear flow and affects how the bacteria will be oriented and packed in the droplet. The cells are oriented along the flow direction near the periphery. They accumulate near the extensional axis of the flow within the interior bulk region, resulting in an emergent anisotropy and heterogeneity.

Recently, Dileep et al. [155] studied drying droplets of Escherichia coli to understand the complex interactions underlying the formation of distinct patterns. The interactions tend to be primarily random for the deposited patterns in the central region of the droplet. The bacteria behave as interconnected clusters like 'sticky particles'. However, the dynamics near the droplet edge tend to be that of the non-interacting particles, with the substrate bacteria interactions dominating. However, to comprehend why and how these patterns form, one needs to analyze the cracks and the mechanical instabilities at play. Recently, a landmark paper analyzed different cracks with *Escherichia coli* in combination with motility traits [156]. It is reported that circular cracks are present for wild-type with active swimming, whereas spiral-like cracks form in the case of non-motile strains of *E. coli* bacteria. The authors have used an elastic fracture mechanics framework and poroelasticity concepts to demonstrate how the tensile nature of the radial drying stress determines circular cracks. Furthermore, it links the microscopic swimming behavior of each bacterium and how it demonstrates the collective behavior, mechanical instabilities, and macroscopic pattern formation. Another study involves bacteria suspended in DI water to understand the self-assembly mechanism, leading to unique patterns. It is found to be dependent on the bacteria-bacteria, bacteria-substrate, and bacteria-liquid interactions. It examines that the shear stress and drying-induced stress on pathogenic bacteria in a sessile drying droplet setting alters the viability and infection [157].

These studies opened up two vital research questions on whether the drying microbial droplet patterns are dependent on: (a) chemotactic behavior and/or (b) secretion of materials (lipids, bio-surfactants, etc.) by the bacteria, which are explored in the later studies. Interestingly, in 2014, Thokchom et al. [77] investigated the chemotactic behavior using Particle Image Velocimetry (PIV), quantifying the velocity fields, initial concentrations, and morphological patterns with and without any nutrient (sugar) gradients on the substrates [see Figure 15(I)]. The authors use the sugar crystal as a chemo-attractant for living and dead bacteria. It was found that the presence of sugar did not alter the hydrodynamics or the internal circulated flows but only the motion of bacteria (in the case of the live cells). The fluid advects the dead bacteria flows with and without the chemo-attractant, whereas the live bacteria alter the symmetric convective flows. The movement of the live bacteria towards the chemo-attractant resulted due to the displacement of the sugar crystal (chemoattractant) from the center to the right side of the substrate. In terms of the morphological patterns, the deposits of live and dead bacteria were found near the periphery independent of the presence of the chemo-attractant due to the drop's radial outward flow. The ring-like depositing patterns are similar to many other colloidal systems. Intuitively, a prominent spot in the interior (central) region was observed in the case of live bacteria with the sugar placed on the substrate center. This spot was displaced to the right side when the chemo-attractant was placed on the right side. It confirmed the influence of chemotaxis in the case of the live bacteria (no such shift was observed in dead bacteria, where the ring-like deposits were observed consistently, in agreement with earlier studies [153]).

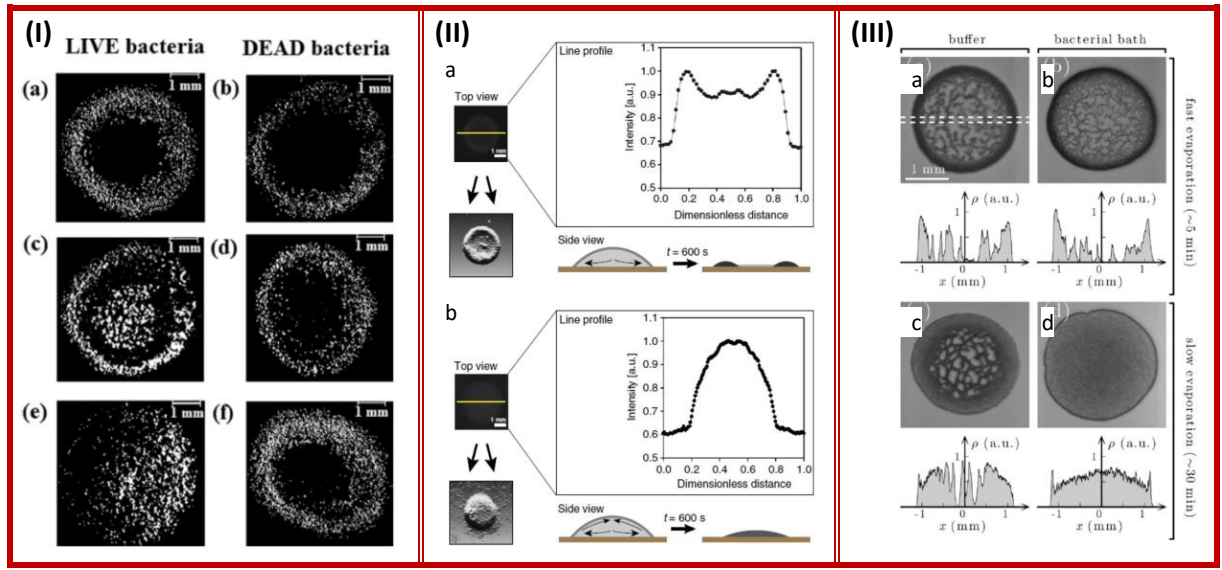

FIG. 15. (I) Patterns of live and dead bacteria for chemotaxis: (a, b) when there is no sugar, (c, d) when sugar on the substrate is at the center, and (e, f) when sugar on the substrate is on the right side, adopted from [77]. (II) Deposition patterns of droplets containing *P. aeruginosa* for a rhamnolipid (bio-surfactant) production deficient mutant (a) and a wild type (b), adopted from [17]. (III) examines the drying rate for patterns of droplets only containing buffer without any bacteria (a) and with bacteria (b) after fast drying and, respectively, (c and d) after slow drying. The plot shows the optical density of the deposit along one droplet's diameter [dashed lines in (a)], adopted from [19].

Sempels et al. [17] examined the role of bio-surfactants produced by bacteria on emerging patterns, specifically, rhamnolipid. Rhamnolipid-producing bacteria form uniform deposi-tion patterns favoring Marangoni flows, whereas the bacteria without rhamnolipid promote outward capillary flow and form a ring-like pattern [see Figure 15(II)]. Surprisingly, during the initial stage of the drying process, the bacteria show swirling flow (vortices). These are carried to the edge first, pulled upwards, and then inward, toward the interior region of the droplet, resulting in larger vortices due to the self-depositing effects of the surfactants produced by these bacteria. In addition, it was noticed that these vortices are generated only above the critical micellar concentration (CMC. For lower than CMC, an oscillatory motion is observed where the capillary flow strength matched that of the Marangoni flow. High bacterial density and the pathogenic nature of some strains make these experiments technically difficult. Different bio-surfactants were added in *E. coli* (not *P. aeruginosa*) at various concentrations to understand whether swirling flow happens due to the bio-surfactants or their physical nature. The presence of vortices in all these systems at the above CMC indicates that the Marangoni flow (due to concentration gradients) opposes the outward capillary flow. These studies successfully uncovered the flow behavior concerning the resulting morphological patterns. However, their practical applicability as a blueprint for detecting characteristic properties of the bacterial species remains unclear.

Besides exploring these above factors, a study by Andac et al. [19] confirms that the deposition patterns and the 'coffee-ring' effect in bacterial systems depend on the drying rate (high or slow), as in the case of the passive (colloidal) particles [see Figure 15(III)]. It also explores the possibilities of bacterial motility on the drying rate. This compares the deposition patterns of the droplet of different buffer solutions with and without motile bacteria. It is observed that the motility is not affected much by the fast-drying rate. Both show uniform deposits, whereas when the drying process gets slower, the patterns differ from each other. This becomes evident when it is found that the rate at which the mass transfer of the bacteria occurs due to the drying process is significantly smaller than the typical speed of motile bacteria in the case of a slow drying rate. This suggests that the self-propelling mechanism of the motile bacteria and the interaction with the environment (different buffer systems) influence the final morphological patterns. Interestingly, a numerical study [158] has been performed to understand the retention and infiltration pathway of the droplets on the leaves. The results have been validated based on available data and the experimental images and confirm that a hydrophilic surface promotes bacterial retention and infiltration. For instance, on a leaf surface, drying-induced flows increase the bacterial concentration around or inside the microstructures of the leaf surface. Larger microstructures and higher drying rates led to higher infiltration. On the other hand, chemotaxis toward nutrients at the leaf surface and random motility decrease the retention and infiltration during the drying process. A recently published paper offers fundamental insights into pattern formation, and links these crack patterns with the underlying mechanical instabilities [156].

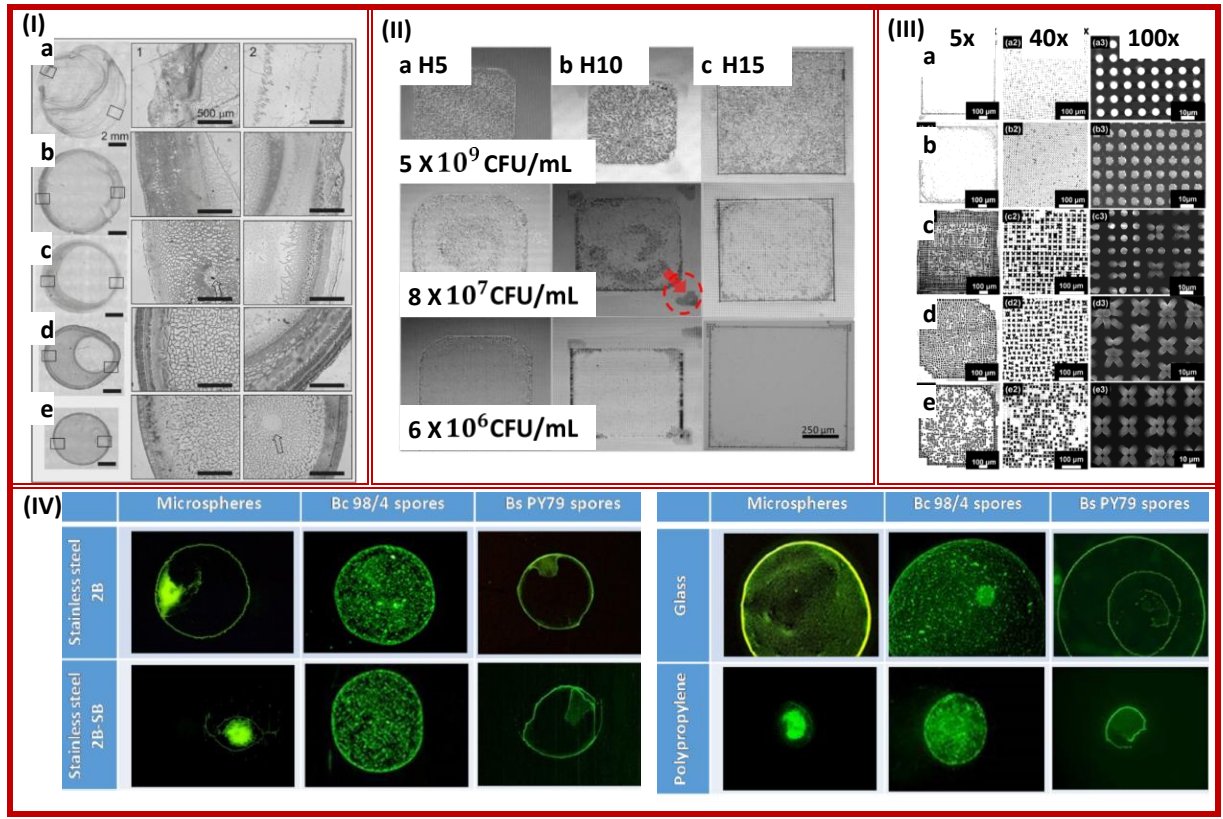

FIG. 16. Bacterial patterns for different substrates: (I) Residues from bacterial suspension on mica surfaces exposed to air: (a) 7 min, (b) 31 min, (c) 95 min, (d) 24 hr, and (e) 92 days, adopted from [159]. (II) patterned surfaces Patterns of *S. epidermidis* droplets containing $5.0 \times 10^9$ CFU mL$^{-1}$ (first row), $8.0 \times 10^7$ CFU mL$^{-1}$ (second row) and $6.3 \times 10^6$ CFU mL$^{-1}$ (third row) over (a) 5 $\mu$m, (b) 10 $\mu$m, and (c) 15 $\mu$m substrates, adopted from [160]. (III) patterned surfaces The different columns correspond to different degrees of magnifications for each strain of bacteria at the end of the drying process: 5$\times$ (left column), 40$\times$ (central column) by using an Optical Microscopy (OM), and 100$\times$ with Scanning Electron Microscopy (SEM) (right column), adopted from [161]. (IV) Deposition patterns of microspheres YG 1.0 $\mu$m and Bacillus spores (Bc 98/4 and Bs PY79) on Stainless steel 2B, 2B-SB, Glass slides, and Polypropylene, adopted from [162].

Looking back, researchers have been interested in also exploring the influence of different substrate properties (conditions) on the patterns formed by the bacterial drying droplets. Baughman et al. [163], in 2010 experimentally studied the bacterial drying droplets deposited on the mica to understand the dependence of their aging process. The patterns formed from a fresh suspension on fresh mica are compared to that of (a) fresh mica but aged bacterial suspension and (b) fresh suspension on aged mica. It was found that both mica aging and bacterial suspension aging affect the patterns, revealing that aging is sensitive to oxygen availability as the suspensions with limited air exposure produced larger residues than those exposed to air during the aging process. However, the residual area is independent of the suspension age when the droplets are deposited on a longer-aged mica. Following up, the authors [159] tuned the morphological patterns by changing the surface wettability, i.e., the contact angle from ∼5 to 20 degrees by timed exposure of the mica surface to the air. The more time the mica surface is exposed to air, the surface wettability decreases (the contact angle increases), and the drop transits from a flat to a taller shape. The shape of the drop modifies, and therefore, the fluid flow during the drying process affects the final morphological patterns [see Figure 16(I)]. For example, ring-like patterns were found near the periphery of the drops for the flat drops due to strong convective and radial flows. In contrast, the accumulation of particles occurs both in the interior and edge of the taller drops. Using Optical Microscopy (OM) and high-resolution Atomic force microscopy (AFM), it is confirmed that the deposition of particles in the interior region changes from multi-rings to cellular (honeycomb) patterns when the drop shape changes to a taller height. In addition, the authors studied whether these patterns emerged solely due to the nature of bacteria or the Minimal Salts Medium (MSM) that was used for preparing the bacterial suspension. It was found that these signature patterns are not observed in MSM-dried residues. However, the deposition in the interior of the droplets increases for both these solutions. In a similar direction, it was observed that the different heights of microstructural patterning on the substrates affect the deposition patterns of the bacteria. The fluid fronts propagate circularly or in a stepwise manner, leading to octagonal or square-shaped deposition patterns [see Figure 16(II)]. In addition, the viability measurements confirm a low deposition of bacteria on the micro-structured surfaces compared to the non-patterned surfaces [160]. Further, experiments with polystyrene beads (of a size similar to bacteria) were conducted to understand the correlation between passive and active particles deposited on the same substrates and drying under the same conditions. Both systems indicate that the low-height patterned substrates can lower bacterial regrowth and biomass attachment on the surface. The same group [161] later designed soft substrates decorated with micro-pillar arrays to understand how the capillary flow in five different bacterial strains is sensitive to these microstructural patterning. The results reveal that the motile bacteria with flagella can stimulate micro-pillar bending, leading to significant distortions and aggregations near pillars forming dimers, trimers, and higher-order clusters. Such deformation could also be thought of as one of the fingerprints. These deformed patterns can be easily seen by the naked eyes [see Figure 16(III)].

It has become clear that the drying rate, substrate properties, and different bacterial strains affect the morphological patterns, but combining these factors was an open question. To fill this gap, Richard et al. [162] examined the overall effects of the bacterial spores by uniting various surface properties, such as topography, hydrophilicity or hydrophobicity, particle properties (size and shape), drying rates, and the deposited patterns. It was found that topography has a weaker influence than hydrophobicity. The spore morphology did not affect the patterns, whereas the hydrophobic spores aggregated to form clusters that quickly get deposited or transported on the surface, leading to a steady distribution of spore clusters [see Figure 16(IV)].

Recently, bacteria deposited patterns are also linked to pathogenesis and virulence [see Figure 17(I)]. The bacterial viability was studied to confirm that the increase in virulence is governed by the flow stress rather than just a function of the nutrition of the media (type and initial concentration of buffer solution). In addition, this study indicates an enhanced virulence and high risk of infection with post-evaporated bacteria on the surfaces [164]. Another study [165] explored the application of these patterns and correlated them with the detection of Antimicrobial Resistance (AMR).

A multi-scale approach was adopted by Hedge et al. [166] to study the spatio-temporal, topological regulation, and bacterial aggregation in drying droplets for the first time. By tweaking flow inside droplets using non-contact vapor-mediated interactions, the authors report the formation of random, multi-scale dendritic, cruciform-shaped deposits when the droplet containing bacteria are dried on glass substrates [see Figure 17(II)]. These deposited patterns; for example, hierarchical dendrite size, orientation, and cruciform-shaped crystals, could be morphologically controlled by tuning the vapor-mediated interactions by placing a droplet of ethanol in the vicinity of this bacterial droplet. The nucleation sites can also be controlled via the preferential transfer of solutes in the droplets; thus, this study finds a new way to control crystal occurrence, growth, and the final topological patterns. This multi-scale approach is not limited to the bacterial solution but can be applied to colloidal drying droplets. Another study uses dark-field microscopy to track any nano-sized particles in different media and shows a potential to be used as one of the *in-situ* techniques for monitoring self-assembled structures and the bottom-up assembly of nano-to-micron-sized particles during the drying process [167].

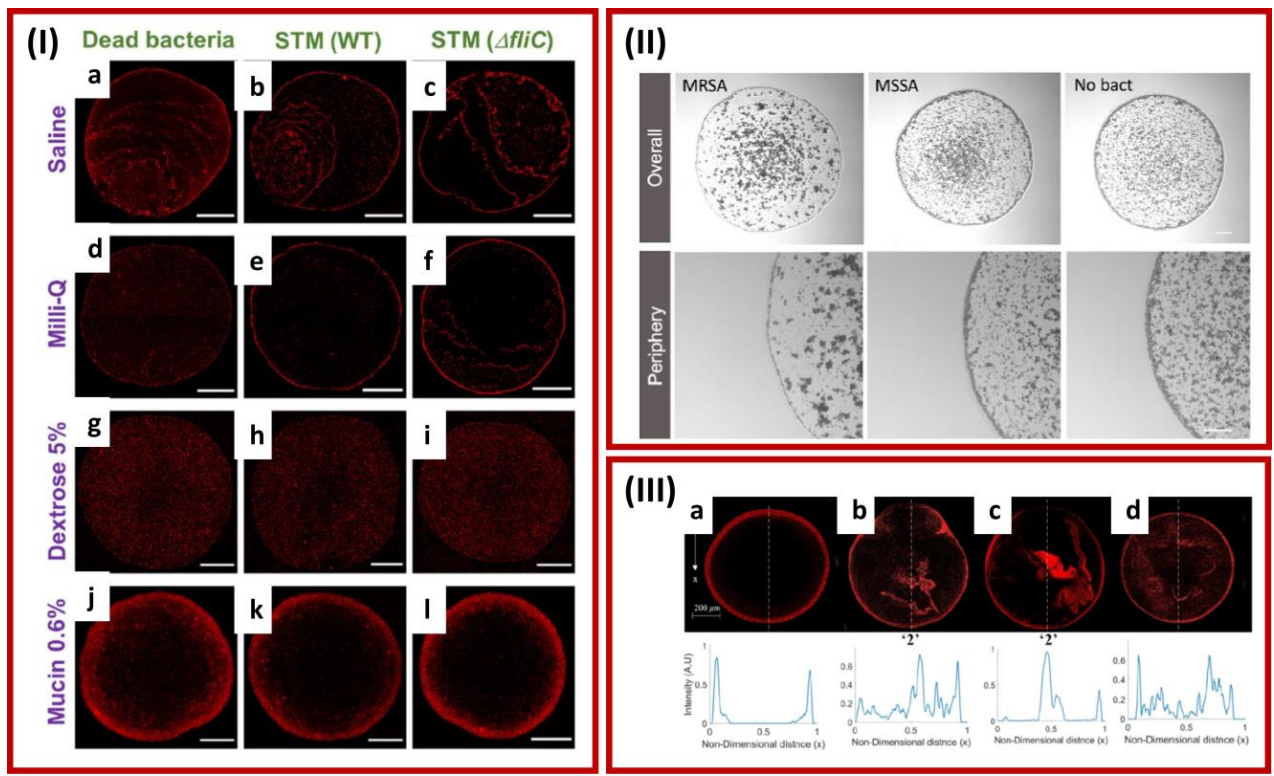

FIG. 17. (I) Patterns for different strains of pathogenic bacteria: dead bacteria, STM (WT), and STM (WT ΔfliC) at different media containing saline, milli-Q, dextrose at 5 wt%, and mucin 0.6 wt%, adopted from [164]. (II) Detection of antibacterial resistance from dried bacterial morphological patterns for MRSA (Methicillin-resistant *S. aureus*) and the drug-sensitive strain MSSA (Methicillin-sensitive *S. aureus*), and control (without any bacteria), adopted from [165]. (III) Emerging patterns in bacteria: (a) in the absence of vapor mediation, and (b-d) in the presence of vapor mediation. Plots in the bottom row correspond to the intensity variation along the center of the droplet, adopted from [166].

Now, going beyond bacteria, can specific patterns emerge in the drying of droplets with other active, living entities? Here we describe studies that investigated bacteriophages, algae, spermatozoa, and nematodes in the drying droplet scenario. Recently, researchers have started studying drying droplets containing pathogenic viruses and bacteriophages that are of current interest due to seasonal influenza and the COVID-19 pandemic, as discussed in the earlier section [131–134]. The study was performed at a fixed initial concentration by varying the culture concentrations to understand the virus viability. It was found that lower culture concentrations suppressed the coffee-ring effect, irrespective of the surface charges of the particles [see Figure 18(I)]. This study also concludes that the coffee rings become more concentrated when there are strong protective effects against the inactivation of these virus aggregations [168]. In another study of bacteriophage, a phase transition from random to oriented structures (chirality) was observed by tuning the rheological properties. The uniaxial or zigzag-like structures are observed by varying the pulling speed of the shear flow and the initial concentration [see Figure 18(II)]. Interestingly, the interplay between the elasticity and hydrodynamic force is found to be responsible during the drying process to obtain such a chiral system [169]. A T4 bacteriophage in the presence of salts is also studied, where the initial concentration of phage and salt ions are found to be the significant factors influencing the fluid flow, particle motion, self-assembled structure, and final morphological patterns [167].

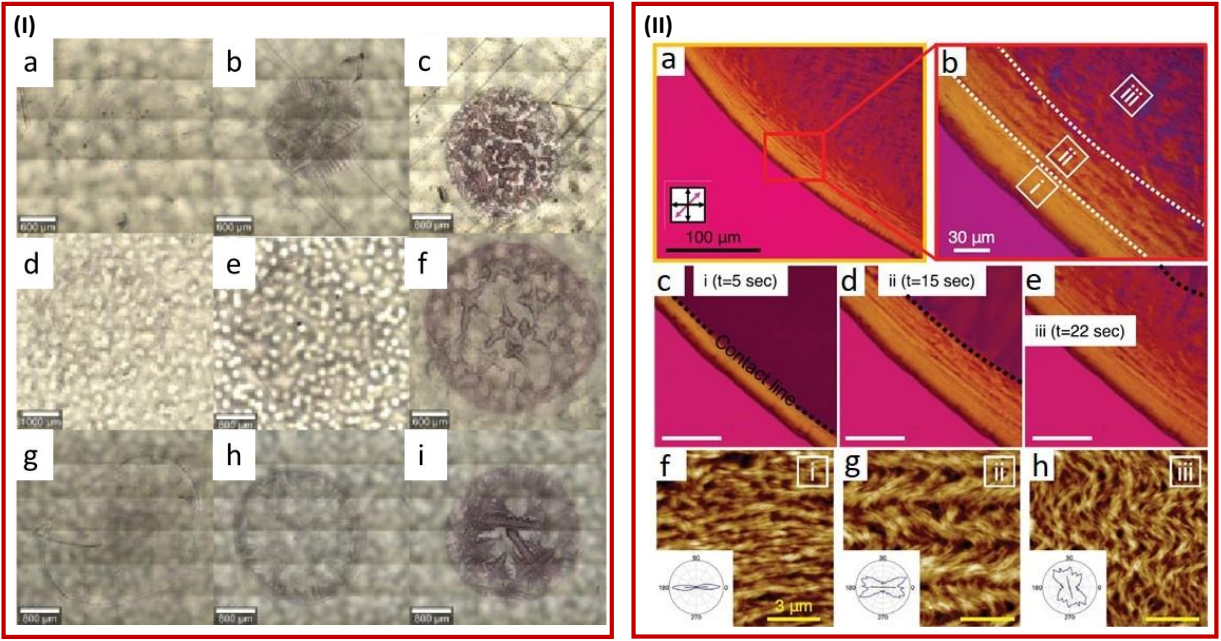

FIG. 18. Dried patterns of bacteriophage: (I) Phi6 droplets for various droplet compositions: (a-c) non-supplemented, (d-f) fixed at 2 wt% FBS, and (g-i) fixed 0.08 wt% BSA at different concentrations of DMEM, adopted from [168], and (II) M13 under Polarizing Optical Microscopy (POM) in (a) and enlarged view in (b). Drying evolution is shown in (c-e) and (f-h) microstructural fibrils measured using AFM. The black, white, and yellow scale bars represent 100, 30, and 3 $\mu$m each, adopted from [169].

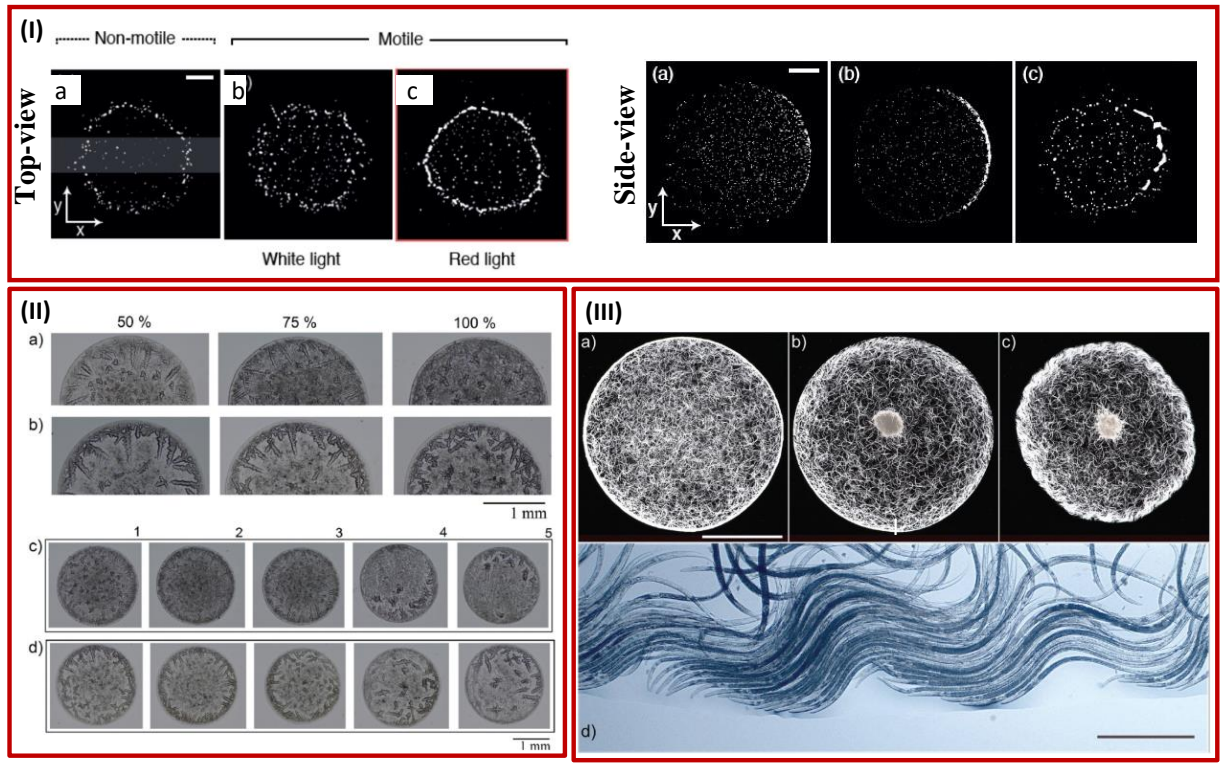

FIG. 19. **(I)** Patterns from algae at different light conditions: (a) non-motile cells, (b) motile cells under white light, and (c) motile cells under red light, captured using the top and side-illumination, adopted from [170]. **(II)** Dried morphology of motile (a) and non-motile (b) spermatozoa, at three concentrations $\varphi$ = 50, 75 and 100%. Patterns formed by the drying droplets containing motile (c) and non-motile spermatozoa (d) at relative concentration $\varphi$ = 75%, adopted from [171]. **(III)** (a–c) Drying evolution of nematode solution at (a) 1 min (random motion), (b) 20 min (percolation of the metachronal wave at the border), and (c) 60 min, with a developed metachronal wave. Scale bar is 5 mm. (d) A view of the metachronal wave similar to that in (c) under a microscope at 4×, with a scale bar of 0.5 mm, adopted from [172].

Another microbial system of interest is the droplet containing photosynthetic algae. It was reported that these algae show a coffee-ring effect and can direct the ring effect spatially depending on the origin of the light source. However, the suppression of the coffee-ring effect, i.e., the uniform deposit, is found in the droplets containing nonmotile algae [see Figure 19(I)]. The deposited patterns are found to be dependent on the light-dependent swimming behavior, unlike other systems studied so far [170]. R´ıos-Ram´ırez et. al [171] examined spermatozoa and found that high concentrations of buffer salts (necessary for the motility of the cells) form complex aggregated distinct structures. These distinct patterns were also found to be influenced by cell motility. During the drying process, initially, the non-motile sperm cells were carried to the droplet periphery, whereas the motile cells showed random motion. At the later stage of the drying process, the concentration of salts increases, leading to non-motile cells in the droplet. Finally, it was noticed that the complex salt aggregates are formed near the droplet periphery; however, more nucleation sites emerged in the droplets containing motile cells [see Figure 19(II)]. In addition, a new image quantification tool, i.e., the polar lacunarity algorithm, was proposed to detect the cell motility in such active systems. Another exciting system scrutinized recently is the nematode [172], which shows promise to be used as a model organism for the experimental investigation of flow behavior in the active drying droplet community. The body oscillation of nematode motion is synchronized [see Figure 19(III)]; however, it cannot be generalized to all the nematodes. For example, *T. aceti* forms traveling metachronal waves, producing a strong fluid flow inside the droplet, whereas *C. elegan*, which has a shorter body size, does not produce any such synchronization during the drying process. This study has attracted the attention of many researchers as nematodes show unique patterns as compared to other active living systems; however, the literature is still at a nascent stage for any generalized conclusion. So far, the microbial studies in the context of drying droplets are limited to the larger droplets as the tiny aerosol-type droplets are technically challenging to image. In 2022, a proof-of-concept was illustrated using two different strains of bacteria as a capillary-clustered aqueous microdroplets dispersed in a thin oil layer. It examined how the drying rate is an essential parameter in deciding the viability of these microorganisms [173].

Since the pattern formation depends on the initial experimental conditions, substrate types, and drying situations, it is important to note each case before comparing the main findings of the available literature. Table IV shows the active bio-colloids' type and strain (bacteria, bacteriophage, nematode, etc.), how their cultures are made non-motile, what kind of substrate is used, and the drying conditions, etc., relevant to the pattern formation. It is to be noted that the drying conditions are tabulated for the main experiments; however, many authors have carried out experiments at elevated temperatures as a control experiment. In addition, only the tools that are used for investigating the drying evolution and the resulting patterns are presented in the imaging techniques. Many authors have measured the contact angle of the droplets and substrate wettabilities that are not listed here. In addition, different microscopy configurations are used by the researchers, such as AFM, SEM, OM, POM, SERS, Confocal Microscopy (CM), Confocal Laser Scanning Microscopy (CLSM), Fluorescence Microscopy (FM), and Polarizing Confocal Microscopy (PCM).

TABLE IV: Summary of the drying droplets of microorganisms with a description of strains, preparations, substrate types, imaging techniques, and references.

| Types | Strains | Preparations of non-motile or dead strains | Substrates | Drying conditions | Imaging techniques | Refs. |
|---|---|---|---|---|---|---|
| Bacteria | *Escherichia coli* DH5$\alpha$ | Cultures were kept for more than a week at RT $(25-30\,^\circ$C) | Thin substrate (mild steel) confined between two parallel glass plates whose inner surfaces were coated with Teflon | Ambient conditions with controlled T = 24$\,^\circ$C, and RH = 30-40% | Visualized with a LED light, and the images were recorded by a CCD camera | [77] |
| Bacteria | *Escherichia coli* RP437 | It is obtained by waiting enough time | Glass slides | Ambient conditions | OM | [19] |
| Bacteria | *Escherichia coli* MTCC443, and DH5R, *Lactobacillus saliVarius* subsp NRRLB 1949 | Cultures were kept in liquid nitrogen for 10 min, followed by immediate thawing to RT | Alcohol-cleaned coverslips | Ambient conditions | OM | [153] |
| Bacteria | *Pseudomonas aeruginosa* PAO1 | | Ruby mica (muscovite) | Ambient conditions | OM, AFM | [159, 163] |
| Bacteria | *(a) Escherichia coli* K-12 [JM109 (DE3)] transformed with eGFP/pRSET, and (b) *Pseudomonas aeruginosa* PA14 wild-type and its rhamnolipid production knock-out mutant | Cultures were exposed to 4% paraformalde-hyde and the antibiotic chlo-ramphenicol at different exposure times | Cleaned coverslips | Ambient conditions with controlled T = $26^\circ \pm 0.5^\circ C$ RH = $38 \pm 1\%$ | CM | [17] |
| Bacteria | *Escherichia coli* (a) wild type: RP437, (b) smooth swimming strain: RP9535, and (c) tumbling strain: RP1616 | | Glass coverslip coated with 0.2% bovine serum albumin solution | Ambient conditions | FM | [18] |

| Bacteria | *Staphylococcus epidermidis* ATTC-12228 | | Epoxy substrates with various height micro-pillars: (a) 5 $\mu$m, (b) 10 $\mu$m, and (c) 15 $\mu$m | Ambient conditions at 21˚ ± 3˚C, and RH = 35 ± 5 % | OM, AFM, SEM | [160] |
|---|---|---|---|---|---|---|
| Bacteria | *Pseudomonas aeruginosa* ATCC-8626, *Escherichia coli* ATCC-10798, *Staphylococcus epidermidis* ATTC-12228, *Bacillus subtilis* ATCC-6051, and *Latilactobacillus sakei* DSMZ-20017 | | Epoxy substrates with various height micropillars: (a) 15 $\mu$m, (b) 22 $\mu$m, (c) 28 $\mu$m, (d) 33 $\mu$m, (e) 38 $\mu$m, and (f) 45 $\mu$m | Ambient conditions at 21˚ ± 3˚C, and RH = 35 ± 5 % | OM, AFM, SEM | [161] |
| Bacteria | *Salmonella Typhimurium* Wild type and ΔfliC | It is obtained by centrifuging at 6000 rpm for 10 min and washed with sterile MilliQ water | Glass slides | Ambient conditions at T = 25−28˚C and RH = 44–48% | CSLM, FM, SEM, AFM | [164] |
| Bacteria | Fluorescently labeled mCherry wild type (WT6) *Salmonella Typhimurium* 14,028 | | Glass slides | Ambient conditions at T = 27° ± 3°C, and RH = 40 ± 5 % | OM | [166] |
| Bacteria | *Escherichia coli* BL-21 DE3 and *Bacillus subtilis* | Propidium iodide was added to the culture | Glass and PDMS slides | Ambient conditions at T = 25°C and RH = 40-50% | FM | [173] |
| Bacteria spores | *Bacillus cereus* CUETM 98/4, and *Bacillus subtilis* PY79 | | Glass slides, Stainless steel (2R, 2B, 2B-SB, 2B-Ugline), Polypropylene | Ambient conditions at T = 20°C | CSLM, SEM | [162] |
| Bacteri-ophage | Phi6 virus was propagated in the bacterial host *Pseudomonas syringae* | | Polypropylene | Ambient conditions at T = 25°C and RH = 25, 50, and 75% | FM, SERS, SEM | [168] |
| Bacteri-ophage | M13 bacteriophage | | The flat silicon wafer substrate | Ambient conditions | AFM, POM, OM, SEM | [169] |

| Algae | *Chlamydomonas reinhardtii* SAG 77.81 | The culture was heated at T = 70˚C | Glass slides | Ambient conditions at T = 20˚C and RH = 50% | PCM | [170] |
|---|---|---|---|---|---|---|
| Sperma-tozoa | Spermatozoa were recovered from the epididymis of male mice CD1 aged 3–5 months | The culture was kept at RT for a day | Glass slides | Ambient conditions at T = 37˚C and RH = 30% | OM, AFM | [171] |
| Nema-tode | *Turbatrix aceti* and *Caenorhabditis elegans* | | Glass slide coated with Rain-X, which contains PDMS as the main ingredient | Ambient conditions at T = 21˚ ± 1˚C, and RH = 15 ± 5 % | OM | [172] |

Therefore, to do any multi-species studies and relate the drying pattern formation, it is recommended that the details of the above aspects are considered. In addition to the number of microorganisms (initial concentration), substrate properties, and environmental conditions, we must consider—(a) strain name and genetic modifications (if any), (b) growth and optimal conditions (temperature, medium, or buffer composition, harvest cycle, etc.), (c) washing methods (filtration or centrifugation, washing steps), (d) procedure to make the microbes non-motile (heating, pH shock, or adding iodide solution). Another point to be noted is that active living droplets are generally multi-component systems with different phenotypes. In the natural setting, microbes do not exist as individual species and always belong in a community. Therefore, to realize the potential of drying microbial droplets, it will be essential to extend the scope of future research to multiple species from natural or built-in communities. For instance, key opportunistic pathogens, including *P. aeruginosa* and *S. aureus*, are frequently present as cooperative or competitive pairs [174] in the environment, food, and body fluids.

## VI. EMERGING CROSS-SCALE AND CROSS-DISCIPLINARY TOOLS AND TECHNIQUES

Many researchers have used simple physical drying processes and studied diverse bio-fluids such as desiccated films, smears, or droplets to understand the pattern formation, often qualitatively relating different diseases like cancer, hyperplasia, fertility, HIV, thalassemia, etc. These patterns are typically evaluated visually (qualitatively) to distinguish and/or identify the diseases. Therefore, many researchers, including us, realize that cross-scale and *in-situ* investigations will be needed to comprehensively understand the signature pattern evolution in the bio-colloidal drying droplets. This investigation should not be limited to the end-point analysis of the drying process. It will also be crucial to include the initial state to the final state of the whole process, i.e., from the dispersed phase (where the particles are still in the solution) to the dried phase (where the components segregate into different complex structures). For instance, the calorimetric measurements of the fluid and the drying droplet method can be combined to gain further insights, as done in [175]. The rheological measurements of the solutions are also crucial [176] to know the visco-elastic properties of the colloidal particles. For example, an aging process is exhibited by laponite clay, and its properties change from liquid to gel-like networks [177]. The rheology and drying process combination might help us better understand the native/initial state of the constituent particles and relate the dried morphological patterns. This cross-scale approach could be differentiated into three different sub-areas or domains which might look independent but are interconnected: (a) different *in-situ* experimental techniques to capture the drying process effectively [see Figure 20(I-VI)], (b) identifying new and robust image processing tools, quantification techniques and predictive modeling via MLAs. Extending the scope of the technical discussion, in this section, we present generalized approaches which could be relevant for both passive and active bio-colloidal and colloidal droplets.

### A. Experimental *in-situ* techniques

In recent years, optical tools and techniques have been vital in driving the *in-situ* research on drying droplets, mainly using laser irradiation and induced heating effects. It is advantageous as it provides a new way of studying the micro-scale phenomena at high-spatiotemporal resolution. Furthermore, it is flexible enough to control different phenomena (mass transportation, Marangoni and capillary flow, etc.) and enables rapid response function capture [184–189]. In 2018, researchers discovered a laser-induced differential evaporation method where the 'coffee-ring' effect can be suppressed without modifying the substrate or using any additives to the liquid. It was achieved by engineering the liquid evaporation profile with a $CO_2$ laser irradiating the apex of the droplets containing fluorescent double-strand DNAs [188]. Along the same line of thought, researchers also used an infrared laser to manipulate crystallization during drying. This method increases the local heating in the droplet, promoting the evaporative crystallization and the resultant Marangoni flow in the droplet [186]. A non-contact light-driven droplet manipulation method has also been introduced for the photothermally active droplets in the drying droplet community. This method induces the Marangoni flow by photothermal heating locally using near-infrared irradiation.

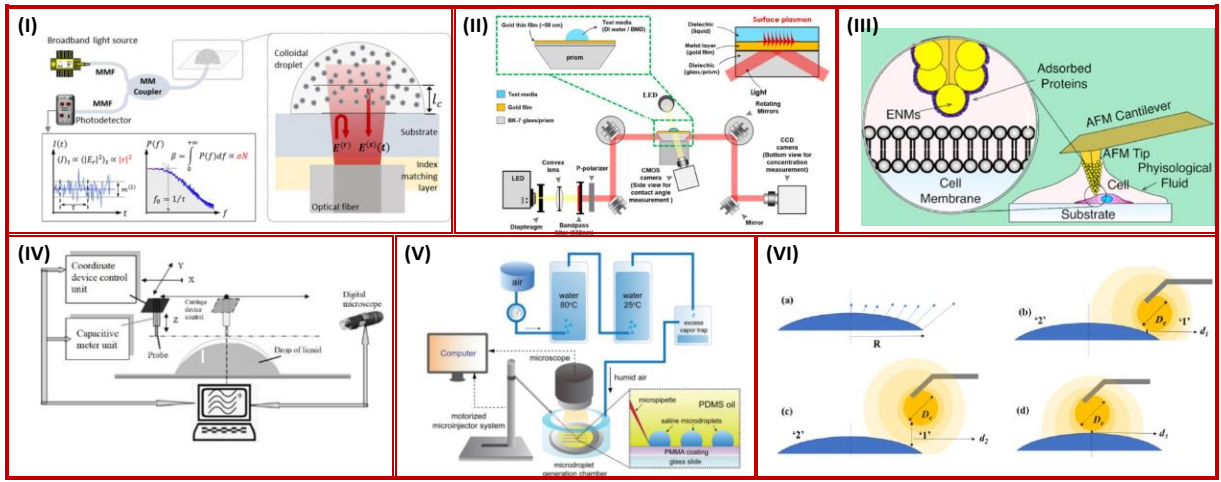

FIG. 20. (I) Schematic of the fiber-based, multimode common path interferometer used to measure the internal dynamics of colloidal droplets, adopted from [178]. (II) The schematic of the experimental setup and principle of SPR, adopted from [179]. (III) Illustration presenting the real-time AFM research strategy, adopted from [180]. (IV) Scheme of the capacitive method for studying drying droplets, adopted from [181]. (V) Schematic diagram of the microdroplet generation system with a humidity control module for induction time measurement via deliquescence-recrystallization cycling, adopted from [182]. (VI) Schematic representation of various cases where an ethanol droplet is moved at different positions in developing vapor-mediated experimental method, adopted from [183].

The droplets slide away from the light, and the direction of motion is influenced by changing the irradiation position [189]. In addition, the researchers have used spatiotemporal coherence-gated light scattering to study the internal dynamics of drying colloidal droplets. The acquired signal originates from a sessile droplet, and the measurement of such a signal is non-contact, non-invasive, and label-free. It also allows capturing the optical and mechanical changes in the droplet during the drying process at the microscopic scale [178]. Another study resolves the issue of having a critical or intrinsic problem, i.e., diffraction limits that prevent us from seeing the sub-micron thin layer formation during the drying process. Therefore, a sophisticated visualization method with a high spatial resolution was adopted for understanding such an adsorption process, i.e., the particle-particle and the particle-substrate interactions using SPR microscopy [179, 190]. In a different study, a simple capacitive method was employed to simultaneously record the height and shape of the droplet during the drying process in the form of a digital signal. This signal can be combined with the visual method to obtain a simultaneous digital image [181]. Similarly, the droplet's physical properties, such as viscosity, composition, surface tension, wettability, and inner structure dynamics, could be measured by the magnitude of AMI signal from a droplet drying on the surface of a quartz resonator oscillating with ultrasound frequency [142]. In the case of saline droplets, nucleation is an important event during the drying process. A novel approach was obtained recently for measuring and analyzing induction times (i.e., how long the droplet takes to nucleate) in sessile arrays of microdroplets using deliquescence or recrystallization cycling. This research methodology offers a promising route in investigating nucleation kinetics and confinement effects and can be applied to pharmaceuticals or other relevant biological systems [182]. In a similar context, researchers have also found a vapor-mediated method by which we can tune the salt-induced dendrite structures formed at the end of the drying process. A droplet of ethanol (volatile liquid) is positioned at different places near the saline droplet. This changes the vapor pressure of the droplet and its surrounding, leading to a more controlled dendrite pattern formation in such saline droplets [183].

Generally, AFM is used as a surface-sensitive technique to characterize dried samples. A real-time AFM would be an innovative step as it will allow us to understand the interaction forces at a nanoscopic scale during the drying process. Recently, AFM has been employed for understanding the particle-cell interactions in physiological media [180], which indicates its potential as a tool to understand the physics of different interactions within the constituent particles in the drying droplet configuration. However, the time-lapse AFM measurements as the droplets dry would be challenging. In contrast, different spectroscopic methods are used extensively for studying disease diagnosis. It is important to combine these two techniques, i.e., spectroscopy and drying droplet [recently known as Drop Coating Deposition Raman Spectroscopy (DCDRS) technique] [191], to understand such a cross-scale complex phenomenon in a better way. Though combining these two techniques is nascent, we would like to bring readers' attention to recent advances. This includes a study on blood plasma and synovial fluids using DCDRS [192], where the chemical compositions of different deposits were investigated and related to the final dried patterns. Recently, the biochemical changes in the tear fluid of a contact lens wearer have been analyzed by this DCDRS technique, which exhibits an alteration of the relative concentrations of proteins and lipids [193]. In this context, Cameron et al. published a review article in 2017 describing how these two combined techniques could impact clinical science [194].

Recently, researchers have developed experimental measuring platforms to study *in-situ* drying processes where the local gas-phase temperature distributions can be quantified. This measuring technique can decouple the energy flows during the drying process to provide information for the liquid-vapor phase change phenomena. This can be applied to different processes, such as condensation, freezing, boiling, etc. [195]. Another *in-situ* measurement was performed by high-resolution neutron imaging to quantify the trace components of the droplet during the drying process. The study is conducted on sessile drying droplets of heavy water and measures the change from usual to heavy water. The *ex-situ* Fourier Transform Infrared (FTIR) supports these dynamic measurements [196]. Furthermore, quantifying the droplet's emerging concentration gradients is vital to understanding the drying-induced stratification patterns. This can be investigated using Confocal Fluorescence Microscopy (CFM) [197].

**B.   Image processing techniques and machine learning algorithms**

Apart from these above-described experimental techniques, developing new image quantification tools and MLAs will be essential to automate the feature identification in the dried droplets. Some bio-fluids show cracks as they dry. Therefore, many researchers have used the crack length, width, etc., as an identification marker [146]. Our group has also established a new processing tool to determine the distance between the consecutive cracks [100]. The texture of the images could also be quantified using FOS and GLCM statistical parameters [95, 198, 199]. The FOS is based on the gray-level distribution of the pixel values without intervening in the interpixel relationships. In contrast, GLCM is second-order statistics, where the parameters are calculated from the spatial relationship between two neighboring pixels. The FOS involves calculating the mean, standard deviation, skewness, and kurtosis of the images, whereas the GLCM involves energy, entropy, inverse difference moment, correlation, etc. [200]. These parameters can also be compared using parametric or non-parametric statistical models. For example, it is shown how these GLCM parameters significantly depend on the horizontal and vertical orientations of the neighboring pixels in an image [198]. Therefore, it cannot be averaged or generalized for all the systems. The lacunarity algorithm (based on the polar symmetry of the circular droplet) and Receiving Operating Curve (ROC) (based on sensitivity and specificity) is also determined from these textures to know how well this distinguishes among deposits quantitatively [171]. In particular, the droplets containing salts exhibit dendrite-like patterns once the salt crystallizes. The length and angle of such dendrites can be calculated, and after that, the fractal dimension could be calculated and used as another image quantitative tool [201]. Most of the bio-fluids show pinning of the droplets to the substrate. And the fluid front moves from the periphery to the central region of the droplet. The radius, velocity, area, etc., could also be used for quantifying the dynamic process during drying. The width of the coffee-ring in the dried states could also provide important information about such droplets [33]. The image processing techniques and statistical methods quantify the features that evolved during the drying process. It includes the detailed procedure to quantify the distance between two consecutive cracks, the texture of the images, and the drying dynamics of the droplets. Statistical methods are also used to support visual interpretation [100].

Many researchers are using PIV to understand flow behavior in the microbial drying droplets, and it could be extended to understand the physical insights of the drying dynamics of the bio-fluids [77]. Recently, the morphological patterns have been analyzed using the frequency distribution method. In this process, the image is divided into small equal intervals and performs a frequency analysis of all distances between all illumined pixels. The frequency analysis captures structural features, whereas the texture analysis (especially skewness) refers to the position of the deposited particles. Another analysis identifies the consecutive structures using the reduction (blur) method. In this process, the number of structures is identified, extending the illuminated points one level outwards, and counting the number of structures. This process of blurring and counting provides information on the complexity. The larger number of structures implies the more heterogeneous the deposit [202]. PIV techniques can be combined with polarization optics, enabling direct visualization and quantitative non-invasive opto-fluidic analysis of the emergent order and structural changes during the evolution of the drying droplets [203–206]. Another exciting image processing technique involves quantitative measurement of changes in the vapor density of the droplet during the drying process using the Schlieren method [207]. It converts the fluid's density into light-intensity information. Various derivative forms of this Schlieren method have been developed, such as the rainbow Schlieren method, background-oriented Schlieren method, etc. Reliable quantitative results using the Schlieren method provide insights into the physics of drying dynamics. Furthermore, infrared thermography with a topographical analysis can be performed on the drying droplet. It provides a highly accurate, fast, and flexible way to extract *in-situ* parameters of Marangoni flow during the drying process in the sheep blood at elevated substrate temperatures [208].

In contrast, standard image processing techniques might not work in this context primarily due to significant variability in specific positions, nature, and characteristics of the features (structures) which evolve during the drying process. To overcome this, machine learning approaches can be used to train on the pattern-recognized network, with robust statistical validation of the geometric differences. In the pre-processing stage, standard image processing libraries, such as thresholding, filtering, etc., remain helpful in identifying crack edges and the network, and a suite of features will be selected based on concepts in geometric network analysis [209] and/or based on Euler characteristics [210]. Such network analytical tools will enable reliable automation pipelines to quantify and characterize features from dried droplets by training suitable MLAs. Other feature-based MLAs include Logistic Regression (LR), Decision Tree (DT), Random Forest (RF), Support Vector Machine (SVM), K-Nearest Neighbors (KNN), and Naive Bayes (NB), to name a few. In contrast, there are several neural-network (NN)-based MLAs, such as Artificial Neural Networks (ANN) and Convolution Neural Networks (CNN). Though MLA is of growing interest in fluid dynamics and diverse drying colloidal systems [211–213], it is relatively recent that this has been used on the drying bio-colloidal droplets [107, 214, 215]. For instance, researchers have optimized the quantification of different images of dried blood droplets, correlating them with different physiological states of cyclists using Principal Component Analysis (PCA), one of the fundamental techniques used for reducing the feature dimensions in MLA [214]. It was also extended to show that MLA and drying droplets can be used as point-of-care diagnostic tools [215].

Furthermore, MLA can detect differences between primary and secondary peptide structures. The feature extraction is obtained from the morphologies generated by the drying droplets, with up to 99% accuracy [107]. However, to the best of our knowledge, such pattern recognition methods were applied to the morphological patterns a decade ago [216], with room for further development, particularly in their accuracy and statistical reliability. Another recent example of MLA applied in the drying droplets of bio-colloids is the identification of milk contamination, coagulation, and decay [212, 213]. The neural network analysis is used for most of these above cases; however, feature-based MLA or a combination of both neural network and feature-based learning would be a plus. It also needs to evaluate which MLA is the best for the small data sets and how we judge their performance levels. Looking at the current scenario of applying MLAs in the drying droplet configuration, we can infer that pattern-recognizing image analyses have shown the potential to overcome variability due to biological or environmental factors. However, applied research on this front is still at a nascent stage.

## VII.   CURRENT CHALLENGES AND PERSPECTIVES

This review highlights the recent progress in drying bio-colloidal droplets and provides a thorough and focused review of the science, applications, and techniques relevant to the field. This section summarizes current challenges and describes our perspectives based on the existing research layout (see Figure 21), providing a concrete tutorial to researchers from different disciplines willing to collaborate toward the next generation of both fundamental and technological science based on drying droplets.

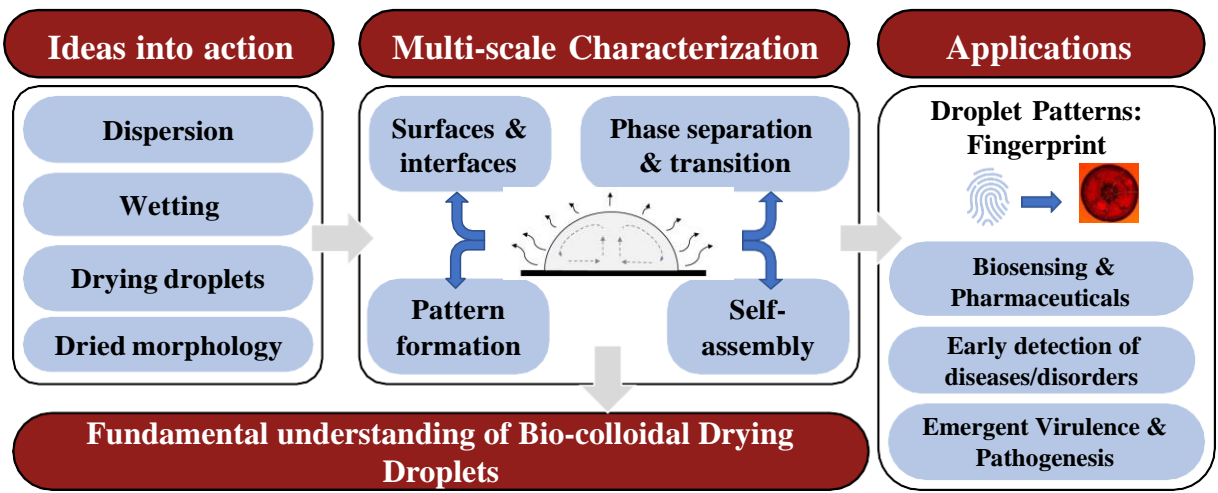

FIG. 21. Presents a perspective on research and applications based on bio-colloidal drying droplets.

## A. Fundamental understanding of bio-colloidal drying droplets

It is known that the pattern formation depends on the type of the solution, initial concentration, substrate on which it rests, environmental conditions (temperature, RH, wind conditions), droplet characteristics (shape, size, methods of deposition, one-to multi-components), etc. Therefore, we need to investigate different stages to understand better the droplet drying process and the consequent pattern formation. Investigations must start from the dispersed states where the bio-colloidal particles are in the solution. Different scattering and spectroscopy techniques (for example, Dynamic Light Scattering (DLS), FTIR [217], SERS [218], etc.), in addition to the polarized visualization of drying droplet methods, can be used as the complementary measurements for a better understanding of the underlying mechanisms. This will allow investigations on the surfaces by varying the substrate properties (patterned vs. smooth in terms of texture, hydrophobic to hydrophilic in terms of wettability, and hard or soft in terms of elasticity). This step is essential to understand how surface wettability differences could influence pattern formation. Simultaneous studies involving drying droplets will shed light on different dynamic states of the drying system. The change in the wettability during the process needs to be characterized by contact angle measurements. However, the tensiometry and surface rheological measurements of human bio-fluids have been very successfully applied to be used as diagnostic tools [219] though there is a necessity for further characterization of how different ions (salts) in the bio-colloids affect interfacial viscoelasticity [220]. The investigations at different stages will help us comprehend the com-ponents' phase separation and flow behavior, hydrodynamic instabilities, and fracturing at different spatio-temporal regimes under different environmental conditions as they dry. This process will also enable us to discriminate different components and how they interact with each other and the substrate. The dried morphologies at various length scales need to be quantified from macro (optical, fluorescence, polarizing microscopy depending on the type of the colloidal solution) to nanoscale by providing structural information about the degree of aggregation with AFM, SEM, and transmission electron microscopy (TEM) techniques.

The cross-scale investigation during the drying process involves different physical mechanisms, such as phase separation, phase transition, self-assembling mechanism, etc. Recently, the drying droplets of bio-colloids have been studied to understand the various phase behavior, the physico-chemical mechanism of phase separation, the condensation of nucleic acids, and the enrichment of ribozyme activities, etc. [221]. A simple bio-mimetic colloidal solution, dextran-polyethylene glycol, deserves attention for examining the phase separation in the drying droplet by tuning their initial concentrations, leading to a change of wettability and compatibility [222]. Recently, a concentration-driven phase transition has been found when diluted human blood droplets and a high concentration of BSA droplets are dried under ambient conditions [31]. Furthermore, to study the interfacial and wetting behavior, a droplet-by-droplet deposition approach is adopted where the incremental layer is build-up from depositing droplets [223]. Another study analyzes dynamic interfacial properties to investigate *in-situ* protein-protein, and protein-particle interactions [224]. Recently, the drying kinetics using the reduction-oxidation (redox) stimulus is also investigated to achieve robust and tunable control of the particle deposition, self-organization, and assembling mechanism of the colloidal particles inside the droplet [225]. In addition, data-driven methods, such as MLA, have been used extensively in fluid dynamics to make real-time predictions of the interfacial phenomena during the drying process [226].

### B. Current challenges of bio-colloidal drying droplets

The first challenge of bio-colloidal drying droplets includes the lack of *in-situ* experimental approaches that can meticulously investigate the phenomena at the microscale. Still, novel tools are needed for characterizing different interactions under dynamic conditions [227]. The investigation comprises feature identification and quantifying the interplay of simulta-neous but different phenomena, including hydrodynamics, flow behavior, colloidal stability, and manifold interactions during the drying process. We have sufficient tools and techniques for quantifying dried samples (section VII A); however, we still lack information across scales as the drying process progresses. A second challenge is tackling the self-assembling mechanisms of multi-components in addition to the drying-driven flows, particle-substrate, and other interactions. These processes happen typically at sub-micron or nanoscales, and experimentally, appropriate visualization techniques are missing. In such cases, we must depend on the numerical and analytical simulation data. Hence, active collaborations between the researchers will be crucial to get the next generation of fundamental and applied research. Publicly available open-source codes, including OpenFoam [228] modules, are valuable for solving the transfer of mass, momentum, and energy equations, including the Navier-Stokes equations (fluid) and advection-diffusion equations (for solutes and heat). These codes can obtain the droplet profile, solute concentrations, and temperatures. Technically, interpretation of the simulated data to uncover different self-assembling mechanisms may require training by experts. Another challenge is the dependence of the emerging patterns on the vast parameter space– which is sometimes hard to initialize experimentally. The fundamental understanding of the fluid dynamics in bio-colloidal drying droplets is in the initial stage. Computational Fluid Dynamics (CFD) might help us deepen this understanding, complementing the experiments, where the experiments will explore the dominant mechanisms (and enable us to discard the less important ones) to be introduced numerically. At the same time, simulations will provide the details (e.g., surface tension gradients) and explore the ample parameter space (statistical details from repeated realizations).

The first current challenge of bio-colloidal drying droplets includes the lack of *in-situ* experimental approaches to easily investigate the phenomena at the microscale, featuring and quantifying the interplay of simultaneous but different phenomena including hydrodynamics, flow behavior, colloidal stability, and manifold interactions during the drying process. We have sufficient tools and techniques for quantifying dried samples (see the section VII A); however, we still lack information across scales as the drying process progresses. A second challenge is tackling the self-assembling mechanisms of multi-components in addition to the drying-driven flows, particle-substrate, and other interactions. These processes happen typically at sub-micron or nanoscales, and experimentally, appropriate visualization techniques are missing. In such cases, we need to depend on the numerical and analytical simulation data; hence, active collaborations between the researchers will be crucial to get the next generation of fundamental and applied research. Publicly available open-source codes, including OpenFoam [228] modules are valuable for solving the transfer of mass, momentum, and energy equations, including the Navier-Stokes equations (fluid) and advection-diffusion equations (for solutes and heat), from which the droplet profile, solute concentrations, and temperatures can be easily obtained. Technically, interpretation of the simulated data to uncover different self-assembling mechanisms may require training by experts. Another challenge is the dependence of the emerging patterns on the huge parameter space– which is sometimes hard to initialize experimentally. The fundamental understanding of the fluid dynamics in bio-colloidal drying droplets is in the very initial stage. Computational Fluid Dynamics (CFD) might help us deepen this understanding, complementing the experiments, where experiments will explore the dominant mechanisms (and enable us to discard the less important ones) to be introduced numerically. At the same time, simulations will provide the details (e.g., surface tension gradients) and explore the large parameter space (statistical details from repeated realizations).

Multi-disciplinarity approach will be a crucial step moving forward, particularly in comparison with previous attempts. Establishing cross-disciplinary teams combining expertise in fluid dynamics, colloids, and health, such as cross-scale experimental tools, image processing techniques, computation and simulation, and MLA, and relating to different characteristics (factors or attributes) will be valuable. Meta-analysis of existing datasets, spanning different expertise, could offer tangible alternatives without immediate collaborative opportunities. In the long run, generalized methodologies to quantify the morphological patterns and minimization of experimental variability can offer reliable datasets that could be linked to potential applications.

### C. Applications of drying droplets in bio-colloidal systems

*1. Pre-diagnosis of diseases, disorders, bio-sensing, and pharmaceutical applications*

An extensive study on saliva and its characteristics [230, 231] has been done for different types of cancer [232–234] and patient attributes [235, 236]. It is evident from the recent preliminary studies that spectroscopic techniques remain in high demand in saliva research for the diagnosis of different types of diseases. This makes a potential bio-fluid for investigating the drying droplet patterns in relation to diseases such as cancer, oral disease, etc. Through this review, we have tried to highlight how the quantitative studies on pattern-recognized drying droplets have the potential to detect different diseases and disorders at various stages, sense and identify different drugs, and track adulterants in any biologically-relevant system. Furthermore, the drying droplet method has the advantage of visualizing the deposits and estimating the trace amounts in such abnormalities. It can potentially be more robust, cost-effective, and easy to implement than other techniques. We can use our current understanding of bio-fluids other than blood and plasma, for example, saliva [230], urine [237], tears [229], sweat, etc.

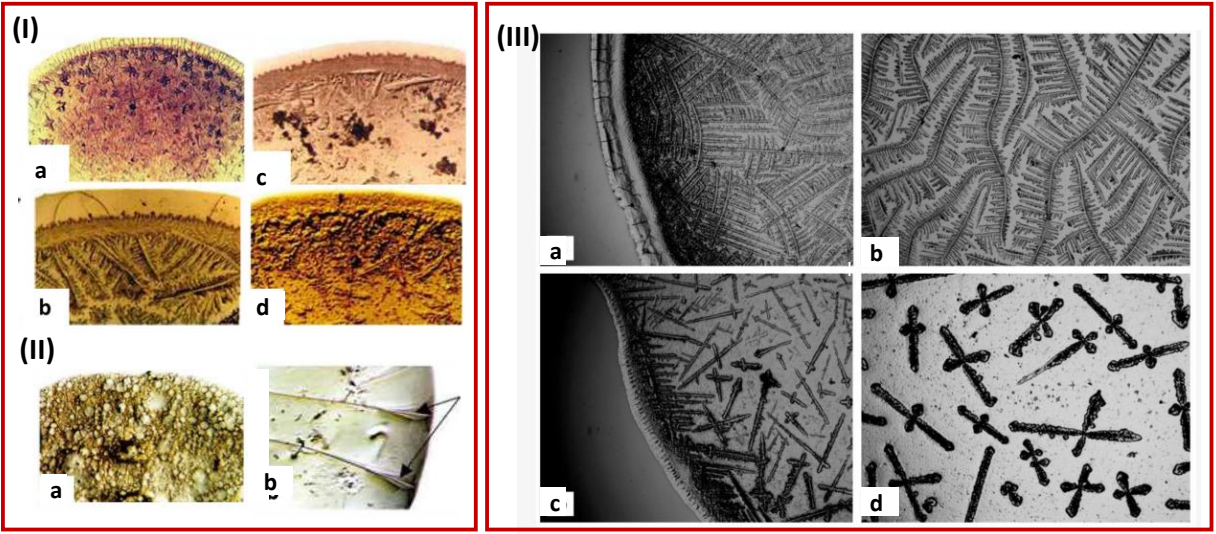

FIG. 22. Dried droplet patterns in different human bio-fluids: (I)(a) healthy tear and tear of patients at (b) early, (c) developed, and (d) terminal stages of glaucoma, (II) (a) healthy synovial fluid and (b) healthy urine with arrows showing leaf-like structures, adopted from [229] and (III) (a-d) healthy saliva with the dendritic growth near the periphery and new crystal nuclei in the central region of the droplet, adopted from [230].

Figure 22(I-II) shows the pattern formation of synovial fluid (fluid in the cavities of synovial joints), urine, tear, and saliva. It also shows the unique morphological patterns for different stages of glaucoma (a disease that damages the eye's optic nerve) in tears [see Figure 22(III)]. An image-corpus could be a good starting point to develop a library of morphological images for the dried samples at each RH, temperature, substrate, etc. (see

Figure 4). Various (appropriate) statistical tools also need to be implemented to identify the key metrics, based on which the standard parameters of interest should be defined for further fundamental studies and to develop a point-of-care diagnostic technique. Furthermore, the droplet drying method will be valuable as it will not require invasive methods for sample collection (e.g., the use of needles for drawing blood), making it highly relevant for infants. The uniqueness of this method is that we can apply the same techniques and protocols for many diseases, including malaria [238], gastrointestinal disorders, COVID-19 related impairments, oral diseases, diet-linked disorders (diabetes) [239], cancer, etc.

In addition to the disease diagnosis, we can quantify the emerging patterns of the drying droplets and relate them to different abnormalities or contamination or a combination of pattern-recognized drying droplets and spectroscopy for industrial and pharmaceutical applications. For instance, this drying-mediated method shows the distinct emerging patterns to detect the presence of contaminants (water, urea, ammonium sulfate, etc.) added to the milk. SERS employed with the drying method detects the presence of urea and ammonium sulfate in milk [240]; even the method is powerful enough to detect trace amounts of added urea (as low as 0.4%) [60]. Another study depicts how distinct patterns can detect medicinal drugs in the solutions [176]. When paracetamol (the drug used for treating high fever and muscle pain) is dried in a simple aqueous solution, it recrystallizes and forms a coffee-ring effect. In contrast, when it is added with different concentrations of chitosan, a polymeric pharmaceutical element often used as a mucoadhesive element and/or antimicrobial agent, it tunes the visco-elasticity of the solution. It suppresses the recrystallization and the coffee-ring effect [176]. In terms of pharmaceutical applications, Carre on et al. [241] studied the quality of different medicinal drugs (methotrexate, ciprofloxacin, clonazepam, and budesonide) as a function of the initial concentration, added NaCl, and substrate temperature. The textural analysis of the morphological patterns provides 95% accuracy in identifying medicines when added in 30% water dilution. It is also found recently that drying-mediated patterns can form uniquely for different homeopathic pharmaceutical preparations [242].

Furthermore, droplet-based bio-sensing methods are attracting many researchers. The pattern-recognized behavior driven by the simple drying process can detect trace amounts of biomolecular concentration with an aptamer-based optical detection based on coffee-ring concentration [243]. A recent review on droplet-based bio-sensing using the drying-induced method can be found in [244]. Recent applications of the drying droplet method for different chemical and biological assays can be found in [245].

Therefore, to establish a point-of-care diagnostic tool, we need to answer the following: (a) Whether these patterns able to differentiate the multiple diseases and comorbidity issues? (b) Is the measurement procedure quantitative? (c) how do we interpret the quantitative results to generate diagnostic reference ranges? Hence, developing robust quantitative methods remain challenging and unresolved. The main barrier is the lack of understanding of how drying patterns evolve and are influenced by the patient's attributes. More advances in new technologies, experimental methods, etc., are needed, though progressing to develop reliable and versatile diagnostics essential to move non-invasive fluids further into mainstream use.

### 2. Emergent virulence, pathogenesis, and anti-microbial resistance

This review article discusses the studies on microbial drying droplets and their pattern formations. So far, investigations have been done on different types of bacteria, surface properties, wetting effects, environmental conditions, etc. However, we believe there is much to learn about bacterial virulence and pathogenesis. This is essential as bacterial virulence, in the context of enhanced pathogenesis and AMR, has long been studied by micro-and molecular biologists in clinical and biomedical settings. The phenomenon of increased drug resistance goes beyond pathogenic bacteria, thus presenting a general trait observed widely across diverse taxa, including the most recent evolution of the COVID-19 variants. Therefore, ecologically relevant biofilm-forming bacterial species in the context of drying droplets might allow us to unravel the links between virulence, phenotypes, and biomechanics. Furthermore, by linking virulence to diverse phenotypes and genotypes in the drying droplet setting, we might account for the additional complexity due to the emergent attributes of drug resistance and thereby pave the way toward the next generation of mitigation and management approaches to solve the AMR challenge. One study recently exhibited that the drying droplet method can test different targets using DNA microbeads. The morphological patterns are different at the macroscale that a smartphone can capture. The dense packing of the microbeads is observed in the central region of the droplet, whereas a thin ring pattern is exhibited by methicillin-resistant *S. aureus*. In addition, this ring is again different for different target bacteria, such as *P. aeruginosa* and *E. coli* [165]. Triggered by the recent COVID-19 pandemic, there is also a growing interest in understanding how the interactions between different species drive them to become pathogenic. It implies that different phenotypic alterations could underpin the mechanism of pathogenesis and, thus, virulence and AMR. In summary, it has a solid potential to deliver a cost and time-effective simple-to-use screening tool for determining AMR, pathogenicity, and overall understanding of virulence and pathogenesis of bacterial strains and their communities.

## VIII.   CONCLUSIONS

In this review, we have highlighted how the physics and engineering of bio-colloidal drying droplets could pave the way for novel and yet thrifty diagnostic toolkits. Successful realization of this bridge between fundamental and applied research will pivot on advancing cross-scale and cross-disciplinary experimental approaches, enabling visualization and quantification of the *in-situ* dynamics during the drying process. The applications of high-end microscopy, image processing techniques and machine learning algorithms have been covered in this review. Thus, this article provides a targeted overview to current and future researchers who may work on drying droplets with biological contexts. The relevant description of the bio-colloids and exploration of hierarchical complexity from relatively simple systems (biomolecules including different types of proteins) to complex fluids such as human blood, serum, and plasma, extracted from healthy people and patients makes this review a significant contribution to both bio-colloidal and synthetic colloidal drying droplet communities. An extensive review of the microbial systems in the sessile drying droplet configuration has been presented for the first time. This compact summary of the recent advancements will allow us to bridge the fields of passive and active material systems via the droplet drying context. Phase separation, self-organizing mechanism, and interfacial properties lead to distinct patterns, which could offer valuable visual cues to specific clinical conditions. To meet the challenges and provide fundamental research and applications, cross-exchanges of ideas and knowledge are essential. Scientific communications and synergistic collaborations between researchers from different fields could be achieved. Scientists from diverse domains, including but not limited to physics, chemistry, biology, and clinicians, need to come together to realize the potential of bio-colloidal drying systems. New findings in the bio-colloidal drying droplets are expected in the next few years, with the possibility of exploring new questions and challenges and driving the sessile droplet community to a broader audience interested in the fundamental and applied prospects of this exciting field.

### ACKNOWLEDGEMENTS

This work is supported by the Department of Physics at WPI, USA, and the University of Warwick, UK. Leverhulme Trust partially funded it (Grant No. RPG-2018345). Our sincere thanks are due to Prof. A. Sarkar, Professor of the School of Food Science & Nutrition, University of Leeds, UK, for providing comments on the initial draft. A. Sengupta thanks the Luxembourg National Research Fund's ATTRACT Investigator Grant (Grant no. A17/MS/11572821/MBRACE) and private donations from the CINVEN Foundation (Covid-19 Fast Track project) for supporting this work.

### AUTHOR CONTRIBUTIONS

Conceived and designed the analysis: A.P and A.G, data collection and compilation: A.P, contributed data or analysis tools: A.P, A.G, and A.S, performed the analysis: A.P, A.G and A.S, wrote the paper (initial draft): A.P, revision and editing the final draft: A.P, A.G., and A.S, and overall supervision: A.P.

---